%% file: Organization.tex
\documentclass[%
 reprint,
superscriptaddress,
preprintnumbers,
nofootinbib,
 amsmath,amssymb,
 aps,
 prd
]{revtex4-2}

\usepackage{amsmath,amssymb}
\usepackage[american]{babel}
\usepackage{microtype}
\usepackage{color}
\usepackage{graphicx}
\usepackage{hyperref}
\hypersetup{colorlinks, citecolor=blue, linkcolor=black, urlcolor=blue}
\usepackage{bm}
\usepackage{hyperref}
\usepackage{relsize}
\usepackage{enumerate}
\usepackage[colorinlistoftodos]{todonotes}

\newcommand{\private}[1]{}
\newcommand{\comment}[1]{}

\newcommand{\be}{\begin{equation}}
\newcommand{\ee}{\end{equation}}
\newcommand{\bea}{\begin{eqnarray}}
\newcommand{\eea}{\end{eqnarray}}

\newcommand{\ket}[1]{|#1\rangle}

\newcommand{\rog}[0]{\rho_{\gamma}}
\newcommand{\vp}[0]{\langle v_{\perp} \rangle}
\newcommand{\eh}[0]{E}
\newcommand{\ehmax}[0]{E_{h\wedge}}
\newcommand{\ehmin}[0]{E_{h\vee}}
\newcommand{\ed}[0]{\mathcal{E}}
\newcommand{\sth}[1]{\makebox[0pt]{\hss#1\hss}}
\newcommand{\mydiamond}{%
  \sbox0{$\lozenge$}%
  \usebox0\kern-.5\wd0\sth{\raisebox{.1ex}{\scalebox{.7}[1]{$-$}}}\kern.5\wd0%
  }
\newcommand{\echar}[0]{E_{\mydiamond}(h)}

\begin{document}

\title{Thermodynamics of ``Continuous Spin'' photons}

\author{Philip Schuster}
  \email{schuster@slac.stanford.edu}
  \affiliation{SLAC National Accelerator Laboratory, 2575 Sand Hill Road, Menlo Park, CA 94025, USA}
\author{Gowri Sundaresan}
 \email{gowrisun@stanford.edu}
 \affiliation{SLAC National Accelerator Laboratory, 2575 Sand Hill Road, Menlo Park, CA 94025, USA}
 \affiliation {Department of Physics, Stanford University, Stanford, CA, 94305, USA}
\author{Natalia Toro}
 \email{ntoro@slac.stanford.edu}
\affiliation{SLAC National Accelerator Laboratory, 2575 Sand Hill Road, Menlo Park, CA 94025, USA}

\date{\today}

\begin{abstract}

Special relativity allows massless particles to have states of different integer (or half-integer) helicities that mix under boosts, much like the spin-states of a massive particle. Such massless particles are known as “continuous spin” particles (CSPs), a term coined by Wigner, and they are notable for their infinite tower of spin polarizations. The mixing under boosts is controlled by a spin-scale $\rho$ with units of momentum. Normally, we assume $\rho=0$. The interactions of CSPs are known to satisfy certain simple properties, one of which is that the $\rho\rightarrow 0$ limit \emph{generically} recovers familiar interactions of massless scalars, photons, or gravitons, with all other polarizations decoupling in this limit. Thus, one can ask if the photon of the Standard Model is a CSP at small but non-zero $\rho$. One concern about this possibility -- originally raised by Wigner -- is that the infinite tower of polarizations could pose problems for thermodynamics. To address this question, we study the thermal evolution of a CSP photon gas coupled to isothermal matter, across CSP helicity modes and phase space.  We find that the structure of the interactions dictated by Lorentz symmetry implies well behaved thermodynamics. When the CSP photon's interactions to charged matter are turned on, the primary $h=\pm 1$ helicity modes thermalize quickly, while the other modes require increasingly long time-scales to thermalize, set by powers of $T/\rho$. In familiar thermal systems, the CSP photon behaves like the QED photon with small $\rho$- and time- dependent corrections to its effective relativistic degrees of freedom. Sizable departures from familiar thermal behavior arise at energy scales comparable to $\rho$ and could have testable experimental consequences. 
\end{abstract}

\maketitle

\input{Main_body}

\begin{acknowledgments}
We thank Javier Acevedo, Lance Dixon, Saarik Kalia, Aidan Reilly and Kevin Zhou for useful discussions over the course of this work. The authors were supported by the U.S. Department of Energy under contract number DE-AC02-76SF00515 at SLAC.
\end{acknowledgments}

\newpage

\appendix
\input{Appendices}

\newpage
\bibliography{References}

\end{document}

%% file: Main_body.tex
    \section{\label{Section_Introduction}Introduction}

At a purely kinematic level, Lorentz symmetry allows massless particles to have states of different integer (or half-integer) helicities that mix under boosts much like the spin-states of a massive particle ~\cite{Wigner:1939cj}.  Such ``continuous spin'' particles (CSPs) possess a spin-scale $\rho$ (with units of momentum) that characterizes the extent of this mixing; in the limit $\rho\rightarrow 0$, the helicity states are exactly Lorentz-invariant. The possibility that $\rho\neq 0$ has never been experimentally studied. 

Until the work of \cite{Schuster:2013pxj,Schuster:2013vpr}, it was usually assumed that theories of CSPs --- if they exist at all --- are unrelated to more familiar theories and are physically irrelevant. However, the soft factor analysis of interacting CSPs in \cite{Schuster:2013pxj,Schuster:2013vpr} and the field theory works in \cite{Schuster:2014hca,Schuster:2023xqa,Schuster:2023jgc} provide strong evidence that consistent interacting CSP theories exist, and their $\rho\rightarrow 0$ limit \emph{generically} recovers familiar theories of massless scalars, photons, or gravitons. In the recent work of \cite{Schuster:2023jgc}, a formalism for computing scattering amplitudes in QED for nonzero $\rho$ was given, allowing the computation and study of finite $\rho$ corrections to QED. (Also see ~\cite{Schuster:2014xja} for a lower dimensional generalization, ~\cite{Bekaert:2015qkt} for fermionic and \cite{Najafizadeh:2019mun,Najafizadeh:2021dsm} for supersymmetric cases using the ``vector superspace'' formalism. Other formalisms, including constrained metric-like, frame-like, and BRST~\cite{Metsaev:2018lth,Buchbinder:2018yoo,Buchbinder:2024jpt} formulations, have also been used to construct actions for fermionic~\cite{Alkalaev:2018bqe,Buchbinder:2020nxn} and supersymmetric~\cite{Buchbinder:2019esz,Buchbinder:2019kuh,Buchbinder:2019sie,Buchbinder:2022msd} continuous spin fields, as well as those in higher-dimensional~\cite{Zinoviev:2017rnj,Alkalaev:2017hvj,Burdik:2019tzg} and (A)dS~\cite{Metsaev:2016lhs, Metsaev:2017ytk, Metsaev:2017myp, Khabarov:2017lth, Metsaev:2019opn, Metsaev:2021zdg, Buchbinder:2024jpt} spaces. Relations between these formalisms are discussed in Refs.~\cite{Rivelles:2014fsa,Alkalaev:2017hvj,Najafizadeh:2019mun}. For reviews and discussion, see Refs.~\cite{Rivelles:2014fsa,Rivelles:2016rwo,Bekaert:2017khg,Najafizadeh:2017tin}.)

At least for the case of a photon with finite $\rho$, it is now possible to more precisely ask the questions: \emph{How much} does a photon's helicity transform under boosts?  What constraints can be derived on the spin-scale $\rho$ of the photon, and how might a non-zero $\rho$ be observed?

To set the stage, we recall a few basic facts about CSP kinematics and interactions. CSP states of four-momentum $k^\mu$ can always be decomposed into eigenstates of the helicity operator $\hat k\cdot \vec{\bf J}$: 
\begin{equation}
\hat {k}\cdot\vec{\bf J} \ket{k,h} =  h \ket{k,h},\label{helicity3V}
\end{equation}
with integer or half-integer eigenvalue $h$.  The sole change from the familiar case $\rho = 0$ to non-zero $\rho$ is that Lorentz transformations $\Lambda$ transform $\ket{k,h}$ into a superposition of states $\ket{\Lambda k,h'}$ with integer $h'-h$. In the limit $\rho\rightarrow 0$, this superposition becomes simply $\ket{\Lambda k, h}$ times a phase $e^{i h \theta(\Lambda)}$ for a suitable $\theta$.   While a theory of $\rho = 0$ particles can consistently include \emph{only} a single helicity $h$ (or two states with $\pm h$, accounting for CPT symmetry), particles of non-zero $\rho$ must possess an infinite tower of states of \emph{all} integer helicities (or all half-integer helicities, a case we will not consider further here).  This fact suggests, at first glance, that CSP physics must be quite different from that of the familiar theories of helcity 0, 1 and 2 particles.

However, the application of Weinberg's famous soft limit of scattering amplitudes revealed that CSP interactions are necessarily well approximated by familiar gauge theories in the limit of small $\rho$ \cite{Schuster:2013vpr}. Only three types of consistent scattering amplitudes can exist in the soft limit, and each displays a ``helicity correspondence'' at CSP energies large compared to $\rho$. In the first, scalar-like amplitude, the helicity-0 mode is emitted most strongly, with an amplitude that is well approximated by ordinary massless scalar interactions, while all other helicity $\pm h$ modes have emission amplitudes suppressed by $(\rho/E)^h/h!$.  In the other two types of amplitude, photon-like and graviton-like interactions, respectively, the helicity $\pm 1$ ($\pm 2$) modes are emitted most strongly, with amplitudes well approximated by ordinary vector gauge theories (perturbative GR), while other helicity amplitudes are suppressed by $(\rho/E)^{|h-1|}$ ($(\rho/E)^{|h-2|}$.).  At energies much larger than $\rho$, the other helicities must be present in the theory, but induce only small effects. In the more complete formalism of \cite{Schuster:2023jgc}, general scattering amplitudes in QED at finite $\rho$ retain this behavior, and the sum over all polarization states is finite, causing no new divergences in cross sections.  

The mere presence of the tower of modes raises a thermodynamic puzzle first mentioned (in a brief remark) by Wigner \cite{Wigner:1963}: in a theory with CSPs, which have infinitely many polarization states, the vacuum's heat capacity per unit volume is formally infinite. This feature, which appears discontinuous between $\rho=0$ and arbitrarily small non-zero $\rho$, is sometimes taken to preclude the relevance of CSPs to the physical world (and is sometimes conflated with possible inconsistencies with black hole thermodynamics in GR). In this paper, we decisively address Wigner's original concerns regarding CSP thermodynamics. 

The key insight is that any physical system has had only finite time to (approximately) thermalize. Helicity correspondence implies that at high-enough energies, only one CSP mode interacts appreciably with matter, while the others decouple. Hot matter quickly thermalizes the principally interacting mode of a CSP (helicity $\pm 1$ in the photon-like case), but other modes take parametrically longer to thermalize. The formal infinity of polarization modes is never physically relevant, because only a finite number of modes will reach equilibrium in any finite time.  As such, late-time corrections to $\rho=0$ thermodynamics are calculable and the $\rho\to 0$ limit is smooth. Indeed, for temperatures $T\gg\rho$, the effects of non-zero $\rho$ on thermodynamics are parametrically small over exponentially long times. 

\subsection{Summary of results}\label{Subsection_Thermodynamics intro}
This paper clarifies and elaborates on the physical conclusion summarized above, with a focus on QED at finite $\rog$, as defined in \cite{Schuster:2023jgc}, i.e.~a CSP ``photon'' with vector-like couplings to matter. 
The qualitative features of our results would apply equally to scalar- or tensor-like CSPs, but we choose the CSP photon as our primary example because it is the case of greatest physical interest. 

We will use the terms `ordinary photon' and `familiar photon' to refer to the QED photon at $\rog=0$. We will consider basic reactions that produce and absorb on-shell photons in a bath of charged particles, which is precisely what the formalism of \cite{Schuster:2023jgc} is appropriate for. However, for our analysis of elementary thermodynamics, we will use soft CSP momentum limits of the scattering amplitudes. We do this because (a) it is simpler than using full amplitudes, enabling analytic study of quantities that involve sums over all helicities, (b) it captures the parametric $\rog$- and energy-scaling of reaction rates even when soft factors are not strictly applicable, and (c) as we will see, the leading changes to thermalization when $\rog \neq0$ appear at CSP energies $E \ll T$, where the soft limit is kinematically justified. We provide additional commentary on the applicability of our soft-factor approximation in Section \ref{Section_Synopsis and more}.

In this paper, we explore in detail the effects of non-zero spin scale $\rog$ on the phase-space evolution of a CSP photon gas coupled to a thermal gas of charged particles. We show that for energies much greater than the spin scale $\rog$ (``correspondence domain" of the CSP gas):
\begin{itemize}
    \item If brought into contact with a thermal bath at finite temperature, a CSP photon can be modeled as the familiar photon with small, time- and $\rog$-dependent corrections to its effective number of degrees of freedom.
    \item There is a strong hierarchy in mode thermalization, where all but the $h=\pm 1$ helicity modes take parametrically long to thermalize in the $\rog\rightarrow 0$ limit. The thermalization of the CSP proceeds increasingly slowly with time.
\end{itemize}

At energies much smaller than the spin scale (``deviation domain" of the CSP gas), we find that:
\begin{itemize}
    \item There is a weak hierarchy in mode thermalization, where a large (but finite, ${\cal O}(\rog/E)$) number of modes thermalize comparably as the $h=\pm 1$ modes, but take parametrically longer to do so at progressively lower energies. 
    \item The radiation power spectrum at ultra-low frequencies is significantly enhanced relative to the familiar black body spectrum.
\end{itemize}

At temperatures $T\gg \rog$, these effects collectively generate parametrically small corrections to the total energy density in radiation, even at times exponentially larger than the photon thermalization time.

The remainder of this paper is presented as follows: We will begin our discussion in section \ref{Section_thermal regimes} with the thermodynamic setup. Subsequently, we discuss CSP behavior in the ``correspondence domain" and ``deviation domain" in turn, introducing the key ideas step-by-step in sections \ref{Section:E>rho} and \ref{Section_E<<rho} respectively. Section \ref{Section_Synopsis and more} will serve as an extended synthesis of results, and we will highlight a collection of open questions in \ref{Section_open questions}. 

\section{Setup}\label{Section_thermal regimes}

We start by considering an idealized thermal system: a gas of charged particles held at a constant temperature $T$ and constant number density. The gas can be relativistic or non-relativistic, we assume it is non-degenerate. At times $t<0$, the CSP photon phase space density is taken to be $f_{h}=0$ for all $h$. 

At time $t=0$, we `turn on' the coupling of the charged particles to CSP photons. The interactions of charged matter produce CSP photons, which begin to thermalize. We assume, in keeping with helicity correspondence, that CSP modes do not have appreciable self-interactions and so thermalize predominantly via interactions with the charged scatterers. With this assumption, the Boltzmann equation can be solved exactly to study the thermodynamic evolution of the photon gas (see Appendix \ref{Appendix_boltzmann derivation}). The phase space density for mode $h$ evolves as: 
\begin{equation}\label{eq_phase space density}
    f_{h}(\eh,t)= f_{h}^{(eq)}(\eh)[1-\exp{(-t/\tau_{h}(\eh) )}] 
\end{equation} 
Here, $\tau_{h}(\eh)$ is the `characteristic thermalization time' of mode $h$ at energy $E$, given by: 
\begin{equation} \label{eq_thermalization time}
\begin{split}
    \tau_{h}(\eh)=f_{h}^{(eq)}(\eh) \bigg[& \int d\Pi_{in}f_{in}d\Pi_{out}(2\pi)^{4} \\&\times \delta^{4}(\Sigma p_{in}-\Sigma p_{out+\gamma_{h}})  |\mathcal{M}|^{2}\frac{1}{\eh}\bigg]^{-1}
\end{split}
\end{equation}
Here,`in' denotes the incoming scatterers, $f_{in}$ are their phase space distributions (which can follow any equilibrium statistics), `out' denotes all outgoing particles except CSP mode $\gamma_{h}$,  $d\Pi \equiv \frac{g}{(2\pi)^{3}}\frac{d^{3}p}{2E}$, and there is an implied product over all species. Any averages over initial and final spins of scattering states (besides the photon) as well as symmetry factors are implicit, and we ignore any effects of Bose enhancement or Fermi suppression from the outgoing states.
$f_{h}^{(eq)}(\eh)$ is the equilibrium distribution of mode h, which will follow Bose-Einstein statistics.
\begin{equation} \label{eq_Bose Einstein phase space distribution}
 f_{h}^{(eq)}(\eh)= [\exp(\eh/T)-1]^{-1}   
\end{equation}

Equation (\ref{eq_thermalization time}) is completely general, but going forward, we will work in the soft limit, with 
\begin{equation} \label{eq_amplitude split with soft limit}
 |\mathcal{M}| = |\mathcal{M}|_{0}\; | \Sigma_{i} q_{i}z_{i}F_{h}(\rog z_{i})|   
\end{equation}
where the subscript `0' denotes the underlying scattering process to which we attach soft photons in the charged legs denoted by `i', with momenta $p_{i}$ and carrying charges $q_{i}$. The soft factor in (\ref{eq_amplitude split with soft limit}) follows \cite{Schuster:2013pxj} and uses:
\begin{align}
   z_{i} &\equiv \frac{\epsilon(k).p_{i}}{k.p_{i}} \label{eq_z} \\
   F_{h}(w) &\equiv 2\frac{( J_{h}(|w|) - c \delta_{h0} ) ~ e^{ih ~\text{arg}(w)}}{|w|} \label{eq_formfactor}
\end{align}
where $\epsilon (k)$ is the polarization vector, $J_{h}(w)$ is a Bessel function of the first kind, and $c$ can be any arbitrary constant - due to charge conservation, the contribution from $c\delta_{h0}$ terms will cancel when the sum over charged legs is taken. To study the thermodynamics from a general scattering process, it will be convenient to work with a single charged leg and leave the sum implicit, and for such calculations it is convenient to set $c=1$ at higher energies ($E \gg \rog$) and $c=0$ at low energies ($E \ll \rog$). We will use this scheme for calculations in this paper. Note that in the $\rog \rightarrow 0$ limit, $F_{h}(\rog z_{i}) \rightarrow 1$ for $h = \pm 1$, $F_{h}(\rog z_{i}) \rightarrow 0$ for all other $h$, and we recover the familiar soft limit factorization of the QED photon amplitude \cite{Weinberg:1964ew} in (\ref{eq_amplitude split with soft limit}).

We will use the term `primary modes' to refer to $h = \pm 1$ modes of the CSP photon, `partner modes' to refer to $h = 0$ and $\pm 2$ modes, and `sub-partner modes' to refer to all other helicities. For the discussion going forward, we allow the charged scatterers to be non-relativistic, with a velocity  $v_{\perp}$ transverse to the soft photon (which was set to unity in \ref{Subsection_Thermodynamics intro}). In the non-relativistic limit, $\rog z \approx \frac{\rog v_{\perp}}{\eh}$. All numerical simulations in this paper use a full thermal distribution of $v_{\perp}$, however we use its average value $\vp$ in equations. We also note here that all numerical simulations in the paper use the full Bessel function soft factors without any simplifications, but we will present several equations that use approximate forms of $J_{h}(x)$ in the applicable regimes as an aid to understanding thermodynamic behavior. Allowing for a non-relativistic scatterer, the ``correspondence domain" will refer to $\eh \gg \rog\vp$ and the ``deviation domain" will refer to $\eh \lesssim \rog\vp$.

From equations (\ref{eq_amplitude split with soft limit}), (\ref{eq_z}) and (\ref{eq_formfactor}), it can be seen that when $\rog\vp/ \eh \ll 1$, the primary modes of the CSP photon behave similarly as the QED photon (with corrections discussed in \ref{Section:E>rho}). When $\rog\vp / \eh \gg 1$, the primary modes couple more weakly to charged matter than the QED photon, and much more democratically as other helicites. Intuitively, the difference in behavior in these two energy regimes can be understood as a consequence of $\rog$ having mass dimension 1 - we can expect deviations from familiar behavior at lower energies. 

In the following two sections, we study these two energy regimes in detail. We will proceed slowly, introducing key concepts as we proceed, and bring the big picture together in Section \ref{Section_Synopsis and more}.

\section{\label{Section:E>rho} Behavior at energies above spin scale $\bigl (E\gg \rog\vp \bigr)$: \\ ``Correspondence Domain"} 

In this section, we study the thermal evolution and behavior of the CSP gas at energies greater than the photon's spin scale. We demonstrate that in this energy regime, a CSP photon gas behaves like a thermal gas of the familiar photon, with sub-dominant, finite, calculable corrections from all other helicities. Since we expect the spin scale to be small in nature, the discussion in this section applies to most energy regimes probed in familiar physical systems (that have $T \gg \rog$). Even when the thermal system under study is ultra-cold ($T \ll \rog$), its dominant behavior could be in this energy regime if the charged scatterers are highly non-relativistic ($\vp \ll 1$). 

We will begin by reviewing the characteristic thermalization times in \ref{Subsec_thermalization time:E>rho}, and show that mode thermalization in the correspondence domain follows a strong hierarchy. This sets the stage for our discussion on how the phase space of the CSP photon is populated in \ref{Subsection_number density: E>rho}, with most partner and sub-partner modes only partially thermal, and contributing only small corrections to the relativistic degrees of freedom from the primary modes. We will subsequently discuss the energy density of CSP gas overall in \ref{Subsection_Energy density: E>rho}. (This discussion is possible at this juncture because all deviation domain contributions to CSP number density and energy density are sub-dominant in familiar thermal systems, and we will justify this in Section \ref{Section_E<<rho}.) We calculate the time dependence of evolution in CSP energy density explicitly, and show that the CSP photon gas increases its internal energy progressively slower as time progresses - thus, we demonstrate explicitly that despite having an infinite helicity tower, the CSP photon gas has finite energy at all finite times. We defer a discussion on the full CSP number density and degrees of freedom to Section \ref{Section_Synopsis and more}.

\subsection{Hierarchy in characteristic thermalization times}{\label{Subsec_thermalization time:E>rho}}

Using the equations for characteristic thermalization time introduced in \ref{Section_thermal regimes} ((\ref{eq_thermalization time}), (\ref{eq_amplitude split with soft limit}), (\ref{eq_z}), (\ref{eq_formfactor})), and using the Taylor expansion of Bessel functions of the first kind \cite[\href{http://dlmf.nist.gov/10.2.E2}{(10.2.2)}]{NIST:DLMF} (which is valid for all helicities throughout the correspondence domain), we find that the characteristic thermalization time of mode $h$ of the CSP photon to leading order follows:
\begin{equation}\label{eq_thermalization time ratio: E>rho}
     \frac{\tau_{h}(\eh)}{\tau_{*}(\eh)} \sim \frac{\bigl\langle|z|^{2}\bigr\rangle}{\bigl\langle|z F_{h}(\rog z)|^{2}\bigr\rangle} \; \approx \; 2^{\Tilde{h}} (\Tilde{h}+1)!^{2} \biggl(\frac{\eh}{\rog\vp}\biggr)^{2\Tilde{h}}
\end{equation}
where we define $\Tilde{h} \equiv ||h|-1|$ to capture the mode `distance from primary modes', and we introduce a benchmark thermalization time $\tau_{*}(E)$ - the characteristic thermalization time of an ordinary photon ($\rog=0$) at the energy $E$ we are interested in.\footnote{Note that a proper calculation of the thermal averages in (\ref{eq_thermalization time ratio: E>rho}) in the non-relativistic limit can reduce the factorial scaling from $(\Tilde{h}+1)!^{2}$ to $(\Tilde{h}+1)!$ for some helicities. Such non-relativistic effects are fully included in all numerical simulations and elaborated in Appendix \ref{Appendix_delta} but omitted from the main text of the paper.} 
\begin{equation}\label{eq_tau*E/tau*T}
    \tau_{*}(\eh) \sim \tau_{*}(T) \biggl( \frac{E}{T} \biggr)^{2}
\end{equation}

\begin{figure*}[] 
    \includegraphics[width=0.8\columnwidth]{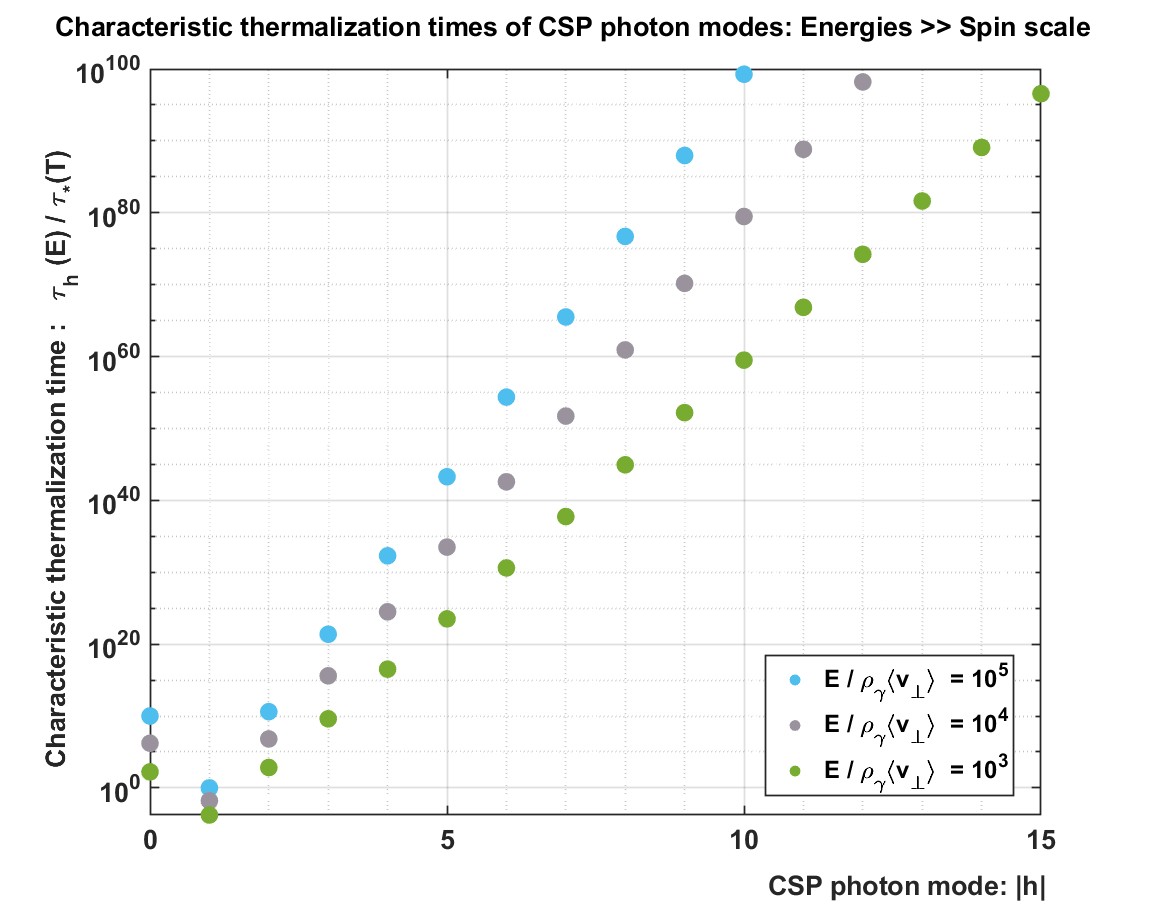} 
    \includegraphics[width=0.8\columnwidth]{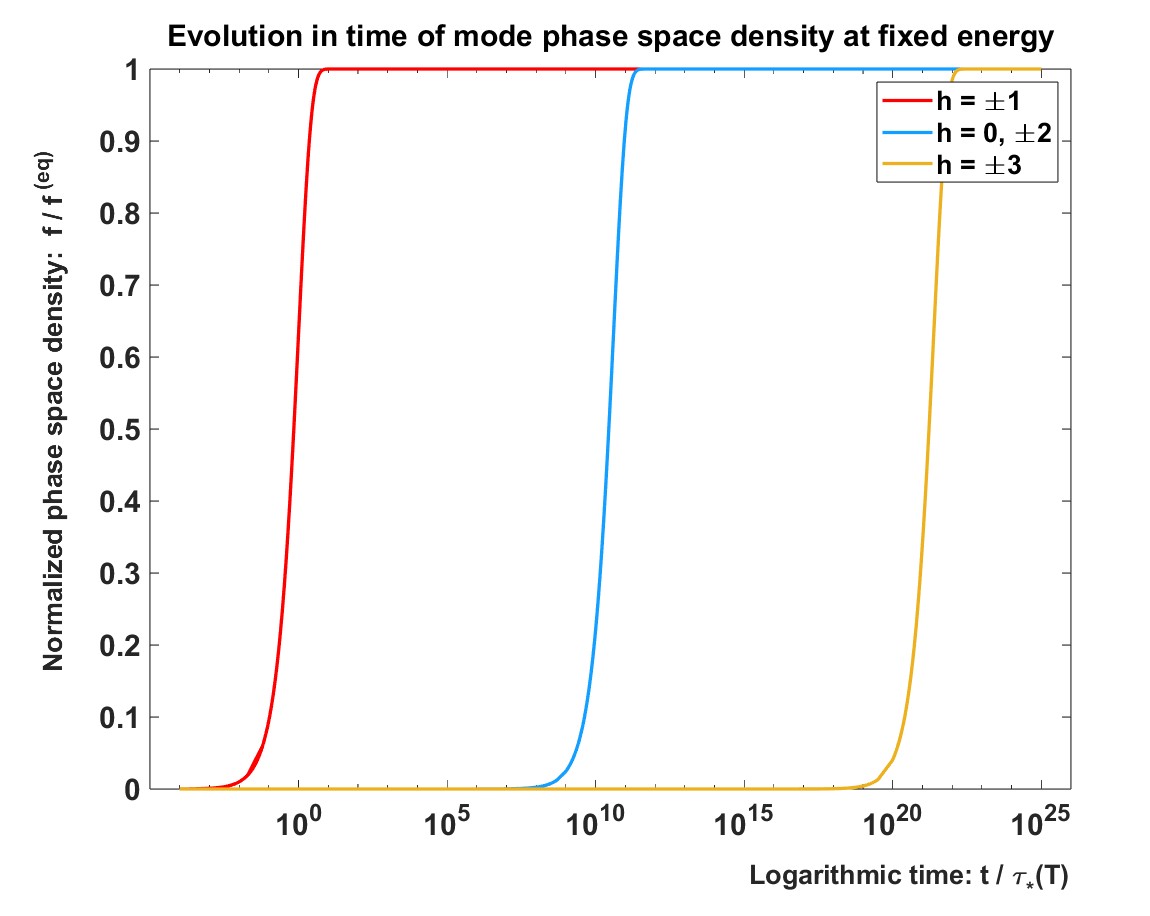} 
    \caption{These figures illustrate that CSP photon modes thermalize with a strong ``double hierarchy" in the correspondence domain ($E \gg \rog\vp$). For these illustrations, we choose $T = 10^{4} \rog$ and $\vp = 0.1$.
    \\ \textbf{Left:} Shows the characteristic thermalization times $\tau_{h}$ of the CSP photon modes at 3 chosen energies $E$, per equation (\ref{eq_thermalization time ratio: E>rho}). The x-axis is the helicity $|h|$. The y-axis is logarithmic in time, and shown in units of characteristic thermalization time $\tau_{*}(T)$ of an ordinary photon undergoing the same thermodynamic process at temperature T. 
    \\ \textbf{Right:} Shows the rate at which primary, partner and $h = \pm 3$ modes are populating their phase space at a given energy (chosen to be E = T) as time evolves, per equations (\ref{eq_phase space density}) and (\ref{eq_thermalization time ratio: E>rho}). The x-axis is logarithmic in time, and shown in units of benchmark thermalization time $\tau_{*}(T)$. The y-axis is the phase space density $f_{h}$ of the mode at the chosen energy, normalized to the equilibrium Bose-Einstein distribution.}
    \label{fig_thermalization times hierarchy: E>rho}
\end{figure*}

For primary modes, $\Tilde{h}=0$, and the ratio in (\ref{eq_thermalization time ratio: E>rho}) reduces to 1 - that is, the primary modes of the CSP photon behave like the familiar photon at leading order in this energy range, with corrections only at $\mathcal{O}(\frac{\rog\vp}{\eh})^{2}$. The other modes, all of which have lower production cross sections, have longer thermalization times. The thermalization of partner modes is suppressed by $\mathcal{O}(\frac{\rog\vp}{\eh})^{2}$ relative to the primary mode. Sub-partner modes thermalize parametrically slower, not only due to higher order suppressions by $(\frac{\rog\vp}{\eh})$, but also due to  suppressions via $ 2^{(\Tilde{h}+1)} (\Tilde{h}+1)!^{2}$, which grows super-exponentially in $h$. Figure \ref{fig_thermalization times hierarchy: E>rho} demonstrates the hierarchy in characteristic thermalization times, and the exponentially longer timescales needed for populating the phase space of the partner and sub-partner modes. 

It is also instructive to review how $\tau_{h}$ of a single mode $h$ depends on energy. It follows from (\ref{eq_thermalization time ratio: E>rho}) and (\ref{eq_tau*E/tau*T}) that $\tau_{h}(E) \propto \eh^{2(\Tilde{h}+1)}$ for all $\eh \gg \rog\vp$. Therefore the low energy phase space close to $\eh \sim \rog\vp$ thermalizes first, whereas the higher energies take longer to thermalize, with a greater suppression at higher helicities.

Thus, the thermalization of a CSP gas at energies above the spin scale is ``doubly hierarchical" (hierarchical in helicity and energy), with higher energy phase space of sub-partner modes getting super-exponentially suppressed. We next present a method to study the effects of this ``double hierarchy".

\subsection{Partial thermalization of modes} \label{Subsection_number density: E>rho}

Due to the strong ``doubly hierarchical" thermalization behavior, a CSP gas as a whole is always partially thermal at any finite time. The phase space density in a mode $h$ at any time $t$ can be obtained by integrating (\ref{eq_phase space density}), and the number density  in the mode follows from it \cite{kolb_and_turner}: 
\begin{align} 
     f_{h}(t) &= \int_0^{\infty} d\eh f_{h}^{(eq)}(\eh)[1-\exp{(-t/\tau_{h}(\eh) )}] \label{eq_phase space density 2}\\
     n_{h}(t) &= \int_0^{\infty} d\eh n_{h}^{(eq)}(\eh)[1-\exp{(-t/\tau_{h}(\eh) )}]
    \label{eq_number density} \\
\text{where } &n_{h}^{(eq)}(\eh) \equiv \frac{1}{2\pi^{2}}f_{h}^{(eq)}(\eh)\eh^{2} \label{eq_equilibirium number density}
\end{align} 
Note that each mode of the CSP carries a single internal degree of freedom. 

At a given time $t$, the low energy phase space for which $\tau_{h} (E) \ll t$ is approximately thermalized. This behavior continues until a maximum energy $\ehmax(t)$ that satisfies the condition $\tau_{h}(\eh)= t$, above which density is well below the thermal value. Thus, the differential phase space density per unit energy at time $t$ (integrand of (\ref{eq_phase space density 2})) can be expressed in terms of its equilibrium value, and the differential number density per unit energy at time $t$ (integrand of (\ref{eq_number density})) has the same form. The latter is\footnote{Note that per (\ref{eq_number density}), at $\eh = \ehmax(t)$, $n_{h}$ has already fallen well below its equilibrium value. (\ref{eq_number density(E_h): E>rho}) thus provides a conservative estimate.}:
\begin{equation} \label{eq_number density(E_h): E>rho}
     n_{h}(\eh,t) \lesssim  
     \begin{cases} n_{h}^{(eq)}(\eh) &\rog \vp \leq \eh \leq \ehmax(t) \\
     n_{h}^{(eq)}(\eh) (\frac{\ehmax(t)}{\eh})^{2(\Tilde{h}+1)} &\eh \geq \ehmax(t)  
 \end{cases} 
\end{equation}
where all variables and parameters are defined as before. Note that a more complete expression for number density will be discussed in Section \ref{Section_Synopsis and more}. We can use (\ref{eq_thermalization time ratio: E>rho}) and (\ref{eq_tau*E/tau*T}) to express $\ehmax(t)$ as:

\begin{equation} \label{eq_Ehmax time dependence: E>rho} 
   \ehmax(t) \sim \biggl[\, \frac{t}{\tau_{*}(T)} \frac{(\rog\vp)^{2\Tilde{h}}}{(\Tilde{h}+1)!^{2}} T^{2}\biggr]\,^{\frac{1}{2(\Tilde{h}+1)}} ~~\text{for all}~ {h}
\end{equation}

For the equilibrium phase space of a mode $h$ to be fully populated by time $t$, its $\ehmax(t)$ has to be $\gg T$. Since $\ehmax(t) \propto t^{\frac{1}{2(\Tilde{h}+1)}}$, it increases progressively slower for higher helicities, and remains close to $\frac{\rog\vp}{|h|}$ for many orders of time after $\tau_{*}(T)$ in the high helicity limit. Differential number density per unit energy at a given time $t$ is illustrated in Figure \ref{fig_number density (E,t): E>rho}. As we expected, most modes are only partially thermal at any given time, and sub-partner modes occupy progressively smaller fractions of their total available phase space.

\begin{figure}
    \includegraphics[width= \columnwidth]{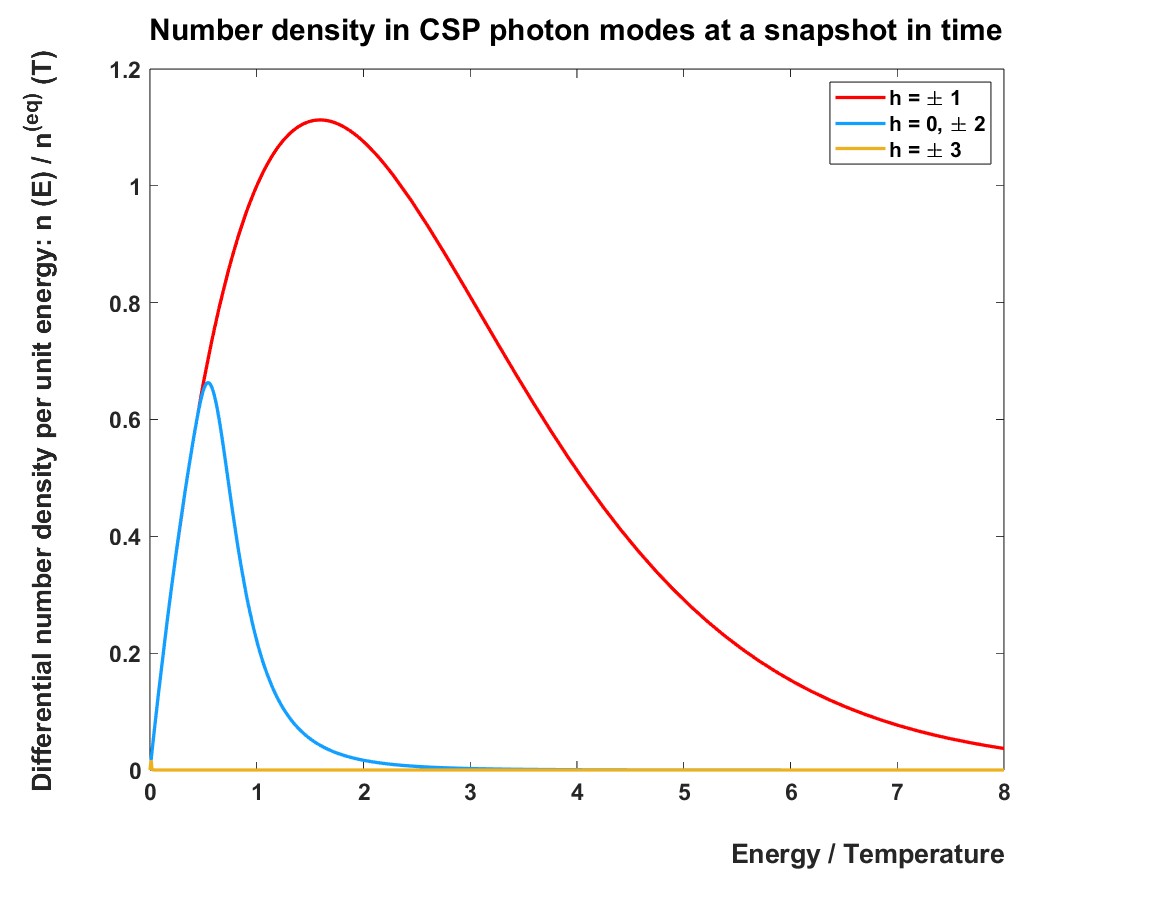}
    \caption{This figure illustrates that at any finite time, the CSP photon gas is only partially thermalized. We show the differential number density per unit energy of CSP modes at a snapshot in time. The x-axis is the energy $\eh$, shown in units of temperature $T$. The y-axis is the mode differential number density per unit energy, normalized to the equilibrium value at temperature T. As shown, the helicity $\pm1$ modes have fully equilibrated, the partner modes have only thermalized their phase space up to $E \sim T$, and the $h = \pm 3$ modes have thermalized a vanishingly small fraction of their phase space (up to $E < 10^{-2}T$). The parameter choices for this illustration are: Temperature $T = 10^{4} \rog$, $\vp = 0.1$, time of snapshot t = $10^{10} \tau_{*}(T)$. 
   }
    \label{fig_number density (E,t): E>rho}
\end{figure}

As mentioned in the section introduction, we will look at CSP energy density overall now, and will return to CSP number density in Section \ref{Section_Synopsis and more}.

\subsection{CSP energy density} \label{Subsection_Energy density: E>rho}

We will demonstrate two critical aspects of the CSP energy density:
\begin{enumerate}
    \item The total energy density is finite at all finite times 
    \item The rate of increase in energy density is inverse in time (once the primary modes are thermal\footref{Footnote_d/dt of primary mode energy density}). That is, the energy density increases more and more slowly as time evolves
\end{enumerate}

The full derivation including simplifying assumptions made is provided in Appendix \ref{Appendix_change in energy density}, but here we present the main results. We will use $\ed$ to denote energy density instead of the more commonly used $\rho$ since we reserve the latter for spin scale.

The energy density in the CSP gas is:
\begin{equation} \label{eq_CSP energy density: all energies}
   \ed_{CSP}(t) = \sum_{h} \int_0^{\infty} d\eh n_{h}(\eh,t) \eh 
\end{equation}

We start with the time rate of change of this.
\begin{align}
    \frac{d}{dt} \ed_{CSP}(t) &\lesssim \sum_{h} \int_{\ehmax(t)}^{\infty} d\eh n_{h}^{(eq)}(\eh)\eh \frac{1}{\tau_{h}(\eh)} \label{eq_d/dt energy density}&\\
    &\approx \Bigg\{ \sum\limits_{h \, : \, \substack{\tau_{h}(T) \geq t \\ h \neq \pm1}} \frac{T^{4}}{2\pi^{2}[2\Tilde{h}-1]} t^{\left( -1+\frac{3}{2(\Tilde{h}+1)} \right)} \nonumber \\
    & \times \left( \frac{\rog^{2}}{\langle p_{\perp}^{2}\rangle} \right)^{\frac{3\Tilde{h}}{2(\Tilde{h}+1)}} \left[\frac{1}{\tau_{*}(T)(\Tilde{h}+1)!^{2}} \right]^{\frac{3}{2(\Tilde{h}+1)}} \Bigg\} \label{eq_d/dt energy density with T/E assumption} \nonumber\\
\end{align} 

All variables and parameters in (\ref{eq_d/dt energy density with T/E assumption}) are as defined before. The conditional sum singles out all the helicities that are still populating their phase space up to $\eh = T$, since at $\eh > T$, energy density increase is exponentially suppressed. We assume the primary modes are thermal and use $\tau_{*}(T)$ as the benchmark time\footnote{When thermalizing, the energy density in primary mode will not change per the form in (\ref{eq_d/dt energy density with T/E assumption}) but instead follows $  \frac{d}{dt} \ed_{*}(t) \propto t^{\frac{1}{2}}$. When a mode is thermalizing, we can generally expect its energy density to change roughly in line with $T \ehmax^{3}(t)$. Refer Appendix \ref{Appendix_change in energy density} for more details. \label{Footnote_d/dt of primary mode energy density}}. 

Even though (\ref{eq_d/dt energy density with T/E assumption}) looks forbidding, its key features are straightforward. First, the rate of change in energy density in all modes is inverse in time, since $\frac{d \ed_{h}}{dt} \propto t^{\bigl( -1+\frac{3}{2(\Tilde{h}+1)} \bigr)}$ for all $\Tilde{h} \neq 0$. This means that even though the mode energy density is increasing with time, it does so more and more slowly. Second, the sum in (\ref{eq_d/dt energy density with T/E assumption}) is convergent, ensured by the $\Tilde{h}^{4}$ factor in the denominator. Lastly, $\rog$ hierarchically controls which helicities participate in the conditional sum, via its control on $\tau_{h}(T)$. 

Due to these factors, the energy density of the CSP as a whole increases very slowly and in a well-controlled manner. At any time, the sum over modes in (\ref{eq_d/dt energy density with T/E assumption}) is dominated by the nearest non-thermal mode (smallest $\Tilde{h}$ participating in the convergent sum), with thermalized modes dropping out of the conditional sum. 
(\ref{eq_CSP energy density: all energies}) and (\ref{eq_d/dt energy density with T/E assumption}) point to a simplified model for the CSP energy density: 

\begin{equation} \label{eq_energy density with time model}
    \ed_{CSP}(t) < \sum\limits_{h \, : \, \tau_{h}(T) < t} \frac{\pi^{2}}{30}T^{4} + \sum\limits_{h \, : \, \tau_{h}(T) \gtrsim t} \int_{0}^{t} dt \frac{d}{dt} \ed_{CSP}(t)
\end{equation}

In (\ref{eq_energy density with time model}), both terms are conditional sums: the first sum is taken over modes that have thermalized by time $t$, whereas the second sum accounts for energy density in modes that are still thermalizing at time $t$. We note here that (\ref{eq_energy density with time model}) overestimates the energy density at any time, since we assume that all the available energy density in a mode i.e., $\frac{\pi^{2}}{30}T^{4}$ \cite{kolb_and_turner} is unlocked when it is thermal up to $\eh=T$. This is a very conservative assumption since only $\approx 3.5 \%$ of total energy density is unlocked by the time any Bose gas has fully thermalized its phase space up to temperature $T$. Additionally, specifically for CSP photon modes, unlocking the additional energy density in the region of phase space above $T$ is parametrically harder since characteristic thermalization times scale as $\eh^{2 \Tilde{h}}$ per (\ref{eq_thermalization time ratio: E>rho}). Nevertheless, we adopt this toy model for its tractability and the aid it provides in building intuition. 

From (\ref{eq_d/dt energy density with T/E assumption}) and (\ref{eq_energy density with time model}), it is easy to see that when a mode is thermalizing, that is, at times $t \lesssim \tau_{h}(T)$, its energy density is increasing as:
\begin{equation} \label{eq_ mode energy density as function of time}
    \ed_{h}(t) < \frac{3}{(\Tilde{h}+1)^{3}}t^{\frac{3}{2(\Tilde{h}+1)}} ~~~\text{for}~ h \,: \, \substack{\tau_{h}(T) \geq t \\ \Tilde{h} \neq 0}    
\end{equation}
Table \ref{table_energy density with time} enlists how 
$\ed_{h}(t)$ is varying for select modes, and compares that with behavior of the primary modes\footref{Footnote_d/dt of primary mode energy density} given by $\ed_{*}(t)$. (\ref{eq_ mode energy density as function of time}) implies that the sum over thermalizing modes in (\ref{eq_energy density with time model}) is convergent at any time. Since there are a finite number of thermal modes at any finite time, the sum over thermalized modes in (\ref{eq_energy density with time model}) is also convergent at any time. Thus, the CSP photon gas always carries finite energy at all finite times. 
\begin{table}[b]
    \centering
    \caption{Upper bounds on increase in mode energy density with time during thermalization (when $\tau_{h}(T) \gtrsim t$) for select modes. $\ed_{*}(t)$ is shown in the first row for comparison.}
    \begin{ruledtabular}
    \begin{tabular}{c|c}
       \textrm{Mode}  & \textrm{Time dependence of $\ed_{h}(t)$ (slower than):}  \\
       \colrule
        $\ed_{*}(t)$ = $\ed_{\pm1}(t)$ & $t^{3/2}$\\
        $\ed_{0,\pm2}(t)$ & $t^{3/4}$\\
        $\ed_{\pm3}(t)$ & $t^{1/2}$\\
        $\ed_{h}(t) ~\textrm{as} ~h \rightarrow \infty$ & Frozen/ Constant\\
    \end{tabular}
    \end{ruledtabular}
    \label{table_energy density with time}
\end{table}

\begin{figure}
    \includegraphics[width= \columnwidth]{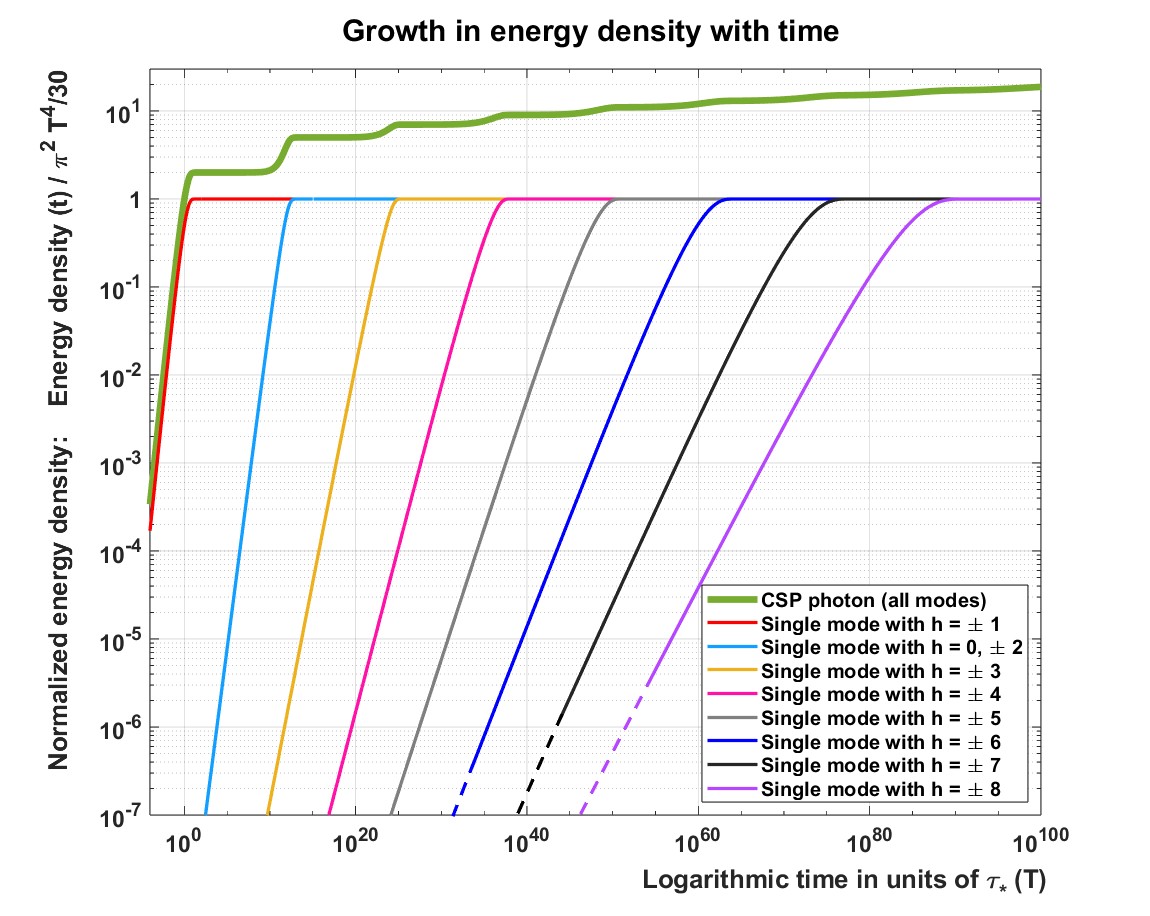}
    \caption{This figure illustrates that the CSP photon gas has finite energy and increases its energy density progressively slowly with time, as indicated by (\ref{eq_d/dt energy density with T/E assumption}) and (\ref{eq_ mode energy density as function of time}). The x-axis is logarithmic in time, and shown in units of benchmark thermalization time $\tau_{*}(T)$ - the time taken for a QED photon undergoing the same thermodynamic process to populate its phase space up to $E = T$. The y-axis is the energy density, normalized to the total energy density in a fully thermalized Bose-Einstein distribution $\frac{\pi^{2}}{30}T^{4}$ \cite{kolb_and_turner}. We show a log-log plot to make the slowing growth rate in energy density manifest. It can be seen that in $10^{100} \tau_{*}(T)$, the CSP photon gas has unlocked $\approx 18.7$ degrees of freedom of the infinite degrees of freedom it has in principle, with a decelerating rate of growth. The parameter choices for this illustration are: Temperature $T = 10^{4} \rog$, and $\vp = 0.1$. Simulations account for mode thermalization behaviors over the full energy range $E = [0, \infty)$, and the dashed lines indicate analytical extensions to simulations.}
    \label{fig_energy density increase with time}
\end{figure}

This is a remarkable result - even when supplied with an isothermal bath that makes infinite energy available to it, the CSP gas takes infinite time to access that energy. This is the crux of why Wigner's infinite heat capacity argument is not physically relevant. Figure \ref{fig_energy density increase with time} illustrates these aspects of CSP energy density. 

\section{\label{Section_E<<rho}Behavior at energies below spin scale $\bigl(\eh \lesssim \rog\vp \bigr)$: ``Deviation Domain" }

As summarized in Section \ref{Subsection_Thermodynamics intro}, the CSP photon gas has a fundamentally different, but still well-controlled behavior at energies $\lesssim \rog\vp$. Since we expect the spin scale $\rog$ to be small, the discussion in this section applies to a very small volume of the overall phase space of the CSP gas in familiar thermodynamic systems (with $T \gg \rog$). Nevertheless, we devote some attention to this since thermal CSP photons in this energy regime exhibit interesting deviations vs. thermal QED photons. In ultra-cold thermal systems (those that have $T \sim \rog\vp$), most of the phase space will be in the deviation domain, and this could have interesting phenomenological implications. Additionally, the well-controlled nature of CSP thermal behavior in the deviation domain is not readily apparent and requires different arguments vs. in the correspondence domain. As before, we begin by studying the characteristic thermalization times. 

\subsection{Weaker hierarchy in \\ characteristic thermalization times}\label{subsection_thermalization times: E<rho}

In this subsection, we show that there is a very weak hierarchy in mode thermalization at energies much smaller than $\rog\vp$. We will see that this weak hierarchy is balanced by the parametrically longer thermalization times as we move down the energy scale and $\eh \rightarrow 0$.

Using the equations for characteristic thermalization time introduced in \ref{Section_thermal regimes} ((\ref{eq_thermalization time}), (\ref{eq_amplitude split with soft limit}), (\ref{eq_z}), (\ref{eq_formfactor})), we find that the characteristic thermalization time of mode $h$ at energy $\eh \ll \rog\vp$ is\footnote{For $\bigl(\frac{\rog\vp}{\eh}\bigr)^{1+\varepsilon} < |h| < \frac{\rog}{\eh}$, Bessel functions cannot be expressed in any simpler form for the entire range of thermal integration in (\ref{eq_thermalization time}). Nevertheless, it can be numerically verified that the Bessel amplitudes fall off rapidly as mode numbers increase. For a relativistic scatterer, the second case in (\ref{eq_thermalization time ratio: E<<rho}) applies for $|h| \gg (\frac{\rog}{\eh})^{1+\varepsilon}$.}:
\begin{align} \label{eq_thermalization time ratio: E<<rho}
    \frac{\tau_{h}(\eh)}{\tau_{*}(\eh)} \sim  
    \begin{cases}
        \bigl(\frac{\rog\vp}{\eh}\bigr)^{3} & |h| \lessapprox \bigl(\frac{\rog\vp}{\eh}\bigr)^{1+\varepsilon} \\ 
         2^{\Tilde{h}} (\Tilde{h}+1)!^{2} \biggl(\frac{\eh}{\rog\vp}\biggr)^{2\Tilde{h}} & |h| > \frac{\rog}{\eh}\\ &\text{and} >\frac{1}{2\vp^{2}}\\
    \end{cases} 
\end{align}
Here, $0 < \varepsilon \ll 1$. $\varepsilon$ has been used to indicate that more modes with $|h| \sim \mathcal{O}\bigl(\frac{\rog\vp}{\eh}\bigr)$ follow similar behavior as the $h \lessapprox \bigl(\frac{\rog\vp}{\eh}\bigr)$ modes. The benchmark thermalization time $\tau_{*}(E)$ still follows (\ref{eq_tau*E/tau*T}). The low helicity case above uses the asymptotic form \cite[\href{http://dlmf.nist.gov/10.17.E3}{(10.17.3)}]{NIST:DLMF} of Bessel functions of the first kind. For the high helicity case, its Taylor expansion \cite[\href{http://dlmf.nist.gov/10.2.E2}{(10.2.2)}]{NIST:DLMF} is valid, just as in the correspondence domain. Equation (\ref{eq_thermalization time ratio: E<<rho}) implies that in the deviation domain, CSP modes follow a fundamentally different thermalization behavior vs. in the correspondence domain. We highlight three key aspects of this behavior.

First, whereas ordinary photons thermalize \emph{more rapidly} at lower energies since $\tau_*(E)$ scales as $E^2$, CSP photon modes with $|h| \lessapprox \bigl(\frac{\rog\vp}{\eh}\bigr)^{1+\varepsilon}$ thermalize \emph{less rapidly} at low energies since $\tau_h(E)$ scales as $E^{-1}$. So, at any time $t$, we can define the lowest energy $\ehmin(t)$ down to which a CSP mode is thermal. We return to this in \ref{Subsection_Characteristic energy} and \ref{subsection_mode thermalization in either direction}. 

Second, in the deviation domain the primary mode is no longer `special'. Instead, as we move to lower energies, an increasing number of modes (those with helicity $\lesssim \frac{\rog\vp}{\eh}$) thermalize on nearly identical timescales, but this timescale also increases as a positive power of $\frac{\rog\vp}{\eh}$. This behavior might seem surprising at first - but it is simply a consequence of lower scattering amplitudes at lower energies, and these scattering amplitudes having a weaker dependence on helicity (up to a limit). As we will see later, this beautiful balance between the number of modes thermalizing and the increasing thermalization time keeps the low energy phase space of the CSP photon well-controlled at all finite times. 

Third, modes with high helicity ($|h| > \frac{\rog}{\eh}$) still take parametrically longer to thermalize - as dicussed above, these modes behave exactly like they do in the correspondence domain, with their thermalization times following (\ref{eq_thermalization time ratio: E>rho}). We provide additional perspectives on this in \ref{subsection_mode thermalization in either direction}.

\begin{figure}
    \includegraphics[width= \columnwidth]{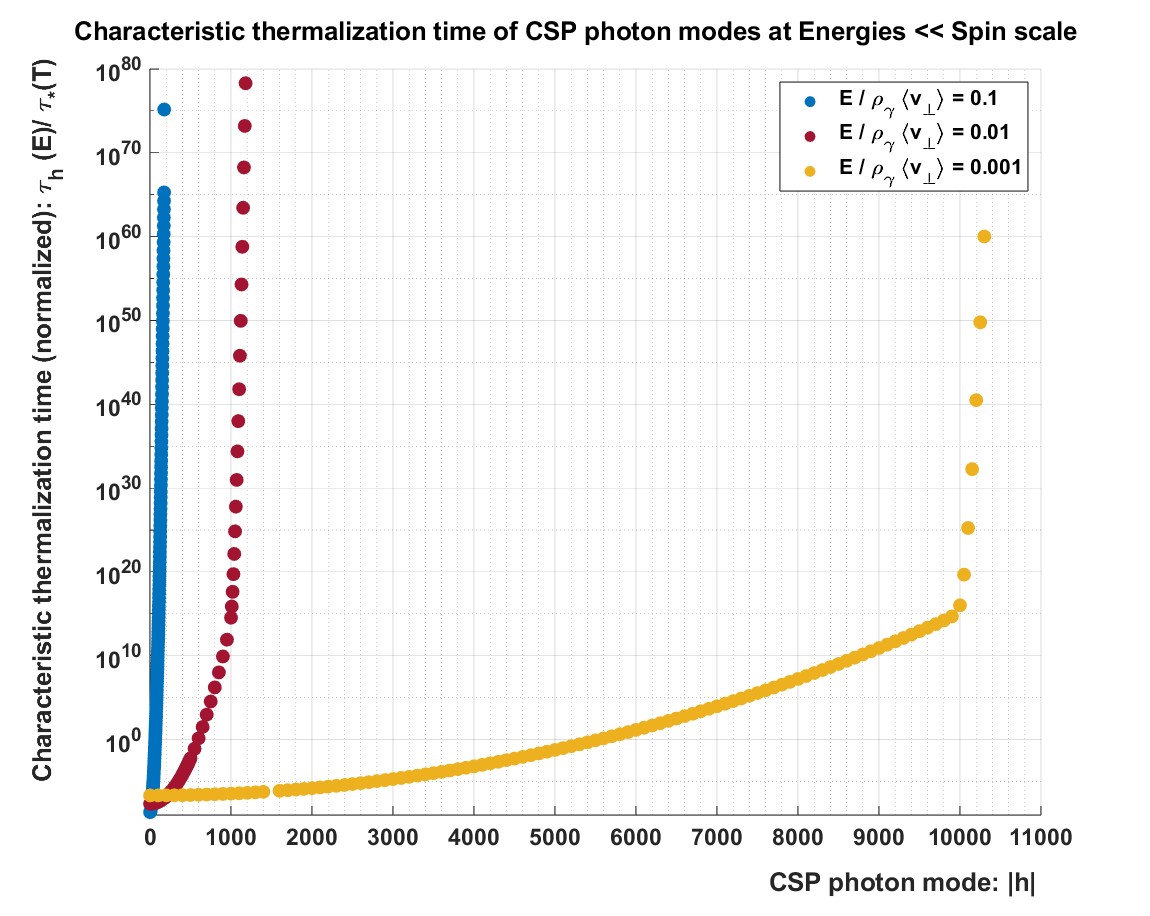}
    \caption{This figure illustrates the controlled thermalization behavior of the CSP photon modes in the deviation domain. It can be seen that the thermalization behavior follows a weak hierarchy, as given by (\ref{eq_thermalization time ratio: E<<rho}), with up to $\mathcal{O}\bigl(\frac{\rog\vp}{\eh}\bigr)^{1+\varepsilon}$ modes identically thermalizing as the primary modes. For the three energies illustrated, it can be seen that $\varepsilon \ll 1$. A strong hierarchy in thermalization behavior, akin to that in correspondence domain, sets in at higher helicities. We illustrate these behaviors at three energies, each separated by an order of magnitude. The x-axis is the helicity $|h|$. The y-axis is logarithmic in time, and shown in units of benchmark thermalization time $\tau_{*}(T)$. The parameter choices for this illustration are: Temperature $T = 10^{4} \rog$ and $\vp = 0.1$.}   
    \label{fig_thermalization times: E<<rho}
\end{figure}

Thus, we still have a hierarchical thermalization behavior, albeit a much weaker one when compared to that in the correspondence domain. The hierarchy in characteristic thermalization times in the deviation domain is illustrated in Figure \ref{fig_thermalization times: E<<rho}. 

We studied the characteristic thermalization times at $E \gg \rog\vp$ using (\ref{eq_thermalization time ratio: E>rho}) and at $E \ll \rog\vp$ using (\ref{eq_thermalization time ratio: E<<rho}). We now talk about an intuitive way to understand mode thermalization behavior in the full energy range, including intermediate energies $E \sim \rog\vp$. 

\subsection{Characteristic energy of a mode} \label{Subsection_Characteristic energy}

We have seen that thermalization time of a given mode increases both at very high energies and at very low energies, with the former controlled by the first term in the Taylor expansion of the Bessel function in (\ref{eq_formfactor}) and the latter by its large-argument asymptotic scaling. In between these two regimes, CSP emission amplitudes are not readily approximated but their behavior can be studied analytically with Bessel functions, and thus can be easily bounded \cite[\href{http://dlmf.nist.gov/10.14}{(10.14.1, 10.14,2)}]{NIST:DLMF}. An important energy scale in the problem is the \emph{characteristic energy} $\echar$ of a given mode, at which that mode thermalizes most rapidly.

\begin{equation} \label{eq_characteristic energy of a mode}
   \echar \approx \frac{\rog\vp}{f^{'}_{h,1}} \approx 
   \begin{cases}
     \frac{\rog\vp}{j^{'}_{h,1}} \sim \frac{\rog\vp}{|h|}  &h \neq 0\\
     \frac{\rog\vp}{1.52} &h=0
   \end{cases}  
\end{equation}

In (\ref{eq_characteristic energy of a mode}), $f^{'}_{h,1}(x)$ refers to the first positive zero of the derivative of $F_{h}(x)$ as defined in (\ref{eq_formfactor}), $j^{'}_{h,1}(x)$ refers to the first positive zero\footnote{The error in approximating $j^{'}_{h,1}(x)$ with $|h|$ decreases as $|h|$ increases. See \cite{DIB:Bessels}} of the derivative of $J_{h}(x)$ \cite{DIB:Bessels}, and $f^{'}_{0,1}(x)$ follows from using $c = 0.5$ in (\ref{eq_formfactor}). Note that $\echar \leq \rog\vp$ for all $h$.

At energies below its characteristic energy, mode thermalization time increases inversely with energy, and at $E \ll \echar$, it follows the equation for the low helicity case in (\ref{eq_thermalization time ratio: E<<rho}). At energies greater than its characteristic energy, mode thermalization time increases with energy, and at $E \gg \echar$, it follows (\ref{eq_thermalization time ratio: E>rho}). At $E=\echar$, the mode thermalization time has a global minimum:

\begin{equation} \label{eq_thermalization time at characteristic energy}
    \frac{\tau_{h}\bigl(\echar\bigr)}{\tau_{*}(T)} \sim \begin{cases}
        \bigl(\frac{\rog\vp}{T}\bigr)^{2} |h| & h \neq 0 \\
        \bigl(\frac{\rog\vp}{T}\bigr)^{2} 1.52 & h = 0
    \end{cases}
\end{equation}

Since a mode thermalizes fastest at its characteristic energy, $\tau_{h}\bigl(\echar\bigr)$ gives the lower bound on $\tau_{h}(\eh)$. Since this lower bound is monotonic in $|h|$, it is straightforward to see that $|h| \rightarrow \infty$ modes need infinite time even to populate the phase space close to their characteristic energy, which is $\approx$ zero.  

\subsection{Mode thermalization behavior} \label{subsection_mode thermalization in either direction}

The notion of characteristic energy (\ref{eq_characteristic energy of a mode}), and the lower bound on thermalization time it provides (\ref{eq_thermalization time at characteristic energy}) allow us to understand mode thermalization behavior throughout the deviation domain, including at intermediate energies $\eh \sim \rog\vp$. At time $t \ll \tau_{h}\bigl(\echar\bigr)$, no region of mode phase space has populated appreciably. At these times,
\begin{equation} \label{eq_number density at early times for all energies}
 n_{h}(\eh,t) \approx   n_{h}^{(eq)}(\eh)\frac{t}{\tau_{h}(\eh)} ~~\text{for all}~~ \eh ~\text{at} ~t \ll \tau_{h}\bigl(\echar\bigr)
\end{equation}
At $t \gtrapprox \tau_{h}\bigl(\echar\bigr)$, the region of phase space close to $\echar$ is mostly populated, but the rest of mode phase space is non-thermal. As time evolves, phase space at energies higher and lower than the characteristic energy thermalize. 

At energies greater than $\echar$, we can define the maximum energy up to which the mode phase space is thermal at a given time: $\ehmax(t)$, introduced and discussed in Section \ref{Subsection_number density: E>rho}. To track how the region of phase space at $\eh \gg \echar$ is populated as time evolves, (\ref{eq_number density(E_h): E>rho}) and the discussion associated with its evolution is valid. It can be verified that at $t= \tau_{h}\bigl(\echar\bigr)$, (\ref{eq_Ehmax time dependence: E>rho}) reduces to $\echar$ in the high helicity limit.  

At energies less than $\echar$, we can define an analogous quantity $\ehmin(t)$: the minimum energy down to which the mode phase space is thermal at a given time. $\ehmin(t)$ tracks how the phase space below $\echar$ is populated as time evolves.
\begin{equation} \label{eq_ehmin(t)}
       \ehmin(t) \sim \frac{(\rog\vp)^3}{T^{2}}\frac{\tau_{*}(T)}{t} ~~~~~~\text{for}~~~ t \gtrapprox \tau_{h}\bigl(\echar\bigr)
\end{equation}
So at energies below characteristic energy, the differential number density per unit energy at time $t$ can be expressed as:
\begin{equation} \label{eq_number density(E_h): E<rho}
     n_{h}(\eh,t) \lesssim   
     \begin{cases} n_{h}^{(eq)}(\eh) &\ehmin(t) \leq \eh \leq \echar \\
     n_{h}^{(eq)}(\eh)\frac{\eh}{\ehmin(t)} &\eh \leq \ehmin(t)  
 \end{cases} 
\end{equation}

We can now gain some intuition for the mode thermalization hierarchies in the deviation domain (\ref{eq_thermalization time ratio: E<<rho}). 

At any given energy $\eh < \rog\vp$, the modes that near-identically thermalize are those that have characteristic energies greater than $\eh$ - so they are all evolving in the phase space region lower than their respective characteristic energies in accordance with (\ref{eq_number density(E_h): E<rho}), with their $\ehmin(t)$ changing as in (\ref{eq_ehmin(t)}) - the helicity independence of (\ref{eq_ehmin(t)}) explains why their evolution is near-identical. As we go down the energy scale, we pass the characteristic energies of more modes per (\ref{eq_characteristic energy of a mode}), and these new modes also start evolving in accordance with (\ref{eq_number density(E_h): E<rho}) and (\ref{eq_ehmin(t)}). 

At any given energy $\eh$, the modes that thermalize with parametrically longer times in (\ref{eq_thermalization time ratio: E<<rho}) are those that have characteristic energies lower than $\eh$, so they are evolving in the phase space region higher than their respective characteristic energies, with thermalization behavior approaching (\ref{eq_thermalization time ratio: E>rho}) at $\eh \gg \echar$. When we get to the correspondence domain, all modes have characteristic energies smaller than the energy we are interested in, and all modes thermalize per (\ref{eq_thermalization time ratio: E>rho}).

\subsection{CSP number density, energy density and \\ spectrum of thermal radiation} \label{subsection_number & energy density: E<rho}

We now discuss three key aspects of the CSP number density and energy density in the deviation domain: 
\begin{enumerate}
    \item The total number density and energy density are both finite at all finite times 
    \item The total energy density in the deviation domain is sub-dominant to the total energy density in the correspondence domain (for systems with $T \gg \rog$)
    \item The thermal radiation from a CSP photon gas is expected to be stronger vs. the standard black body spectrum of a thermal QED gas
\end{enumerate}

First, CSP number density and energy density in the deviation domain are finite since only a finite number of modes contribute to them at all finite times. Those modes for which $\tau_{h}\bigl(\echar\bigr) > t$ would not have populated any portion of their phase space to equilibrium densities, and since $\tau_{h}\bigl(\echar\bigr)$ varies as a positive power of $|h|$, infinite time is needed for the deviation domain of the CSP to fully thermalize. Additionally, the non-equilibrium contributions summed over all CSP modes is also finite since the total number of contributing modes is almost exactly offset by the fractional thermalization at that energy (per (\ref{eq_number density at early times for all energies}) and (\ref{eq_thermalization time ratio: E<<rho})) - so, the low volume of phase space at these energies fully regulates the number density and energy density in the deviation domain at all finite times. 

Second, the total energy density in the region of phase space below $\rog\vp$ remains sub-dominant (assuming $T \gg \rog\vp$), despite many more modes thermalizing simultaneously. It can be shown that\footnote{The total energy density in any mode is most sensitive to its $\ehmax(t)$, and relatively insensitive to its $\ehmin (t)$, so we can get an order of magnitude estimate of the total energy in any mode by tracking only the behavior of its $\ehmax(t)$. This simplifying assumption can be further justified by noting that the total energy contribution to the CSP from thermalization of every mode's phase space region $\ehmin(t) \leq \eh \leq \echar$ is at all times bounded by $\mathcal{O}((\rog\vp)^3 T)$. The total contribution to CSP deviation domain energy from every mode's $\echar \leq \eh \leq \ehmax(t)$ is dominated by the contribution from the highest $\ehmax(t)$ at any time $t$ i.e., the energy at the `boundary region' $E \sim \rog \vp$.} we can get an order of magnitude estimate of the total energy in the deviation domain of the CSP gas using only the energy in the `boundary region' $E \sim \rog \vp$, which can be expressed in the suggestive form: 
\begin{equation} \label{eq_deviation domain energy density}
    \ed_{deviation ~domain} \sim \mathcal{O}\biggl(~\biggl(\frac{\rog\vp}{T} \biggr)^{3} h_{max} T^{4} \biggr)
\end{equation}
where $h_{max}$ is the highest helicity that has populated its phase space at $E = \rog\vp$, given by the mode that saturates the condition $\ehmax(t) = \rog\vp$ in (\ref{eq_Ehmax time dependence: E>rho}): 
\begin{equation} \label{eq_maximum helicity with ehmax = rhovp}
    \frac{t}{\tau_{*}(T)} \biggl(\frac{T}{\rog\vp} \biggr)^{2} \approx 2^{h_{max}} ~ h_{max}!^{2}
\end{equation}
From (\ref{eq_deviation domain energy density}), it is clear that for systems with $\rog \vp \ll T$, the deviation domain of the CSP gas contains a negligible fraction of the total energy in even a single fully thermalized mode of a Bose gas, which is $\mathcal{O}(T^4)$. The only way this can be circumvented is if at a given time, $h_{max}$ can be $> \mathcal{O}\biggl(\frac{T}{\rog\vp} \biggr)^{3}$ \emph{and} those modes which have $\ehmax(t) > \rog \vp$ at that time (which are necessarily smaller helicities than $h_{max}$) have not thermalized much of their correspondence domain. This condition can never be met since the energy scaling of characteristic thermalization time at $E \gg \echar$ is $\tau_{h} \propto \eh^{2(\Tilde{h}+1)}$ whereas the helicity scaling is the much stronger $\tau_{h} \propto (\Tilde{h}+1)^{2(\Tilde{h}+1)}$. Simply put, it is \emph{much} easier for smaller helicities to populate their higher energy phase space than it is for larger helicities to even populate their phase space at energies close to their characteristic energies - at any time, it is impossible for $h_{max}$ to grow large enough to compensate for the energy contained in the correspondence domain of the smaller helicities. 

Third, despite the sub-dominance of the total deviation domain energy of the thermal CSP gas, its black body spectrum shows substantial deviations from that of the familiar thermal QED gas, both in radiated power and spectrum shape. Due to near-identical thermalization of up to $\mathcal{O}\bigl(\frac{\rog\vp}{\eh}\bigr)^{1+\varepsilon}$ modes at energies $\ehmin(t) \leq \eh \leq \rog\vp$, the CSP differential energy density stays nearly linear\footnote{There will be deviations from linearity because: a) $0 < \varepsilon \ll 1$, and b) If we wait long enough, we can expect contributions from the modes that are populating the region of their phase space above their characteristic energies.} down to a time-dependent low energy cutoff. Below this low energy cutoff, no mode has populated its phase space to the equilibrium value, but non-equilibrium contributions of up to $\mathcal{O}\bigl(\frac{\rog\vp}{\eh}\bigr)^{1+\varepsilon}$ modes can still add up per (\ref{eq_number density at early times for all energies}), resulting in a quadratic spectrum shape similar to QED gas in the deep IR, albeit with stronger radiated power\footnote{Small deviations from quadratic form can exist in principle because $0 < \varepsilon \ll 1$. However, as can be seen in Figure \ref{fig_thermalization times: E<<rho}, $\varepsilon$ decreases as we move to lower energies, so these deviations are likely to be negligible.}. Thus the CSP gas radiates stronger than the QED gas at all frequencies in the deviation domain, with exact spectrum shape dependent on allowed thermalization time, the spin scale and temperature. This is in contrast to the QED photon thermal spectrum that remains quadratic in frequency at all $\omega \ll T$. Since the power radiated by the CSP photon at these very low frequencies is spread across modes with suppressed matter interactions, further study is needed to evaluate the detectability of this excess.

\begin{figure}
    \includegraphics[width= \columnwidth]{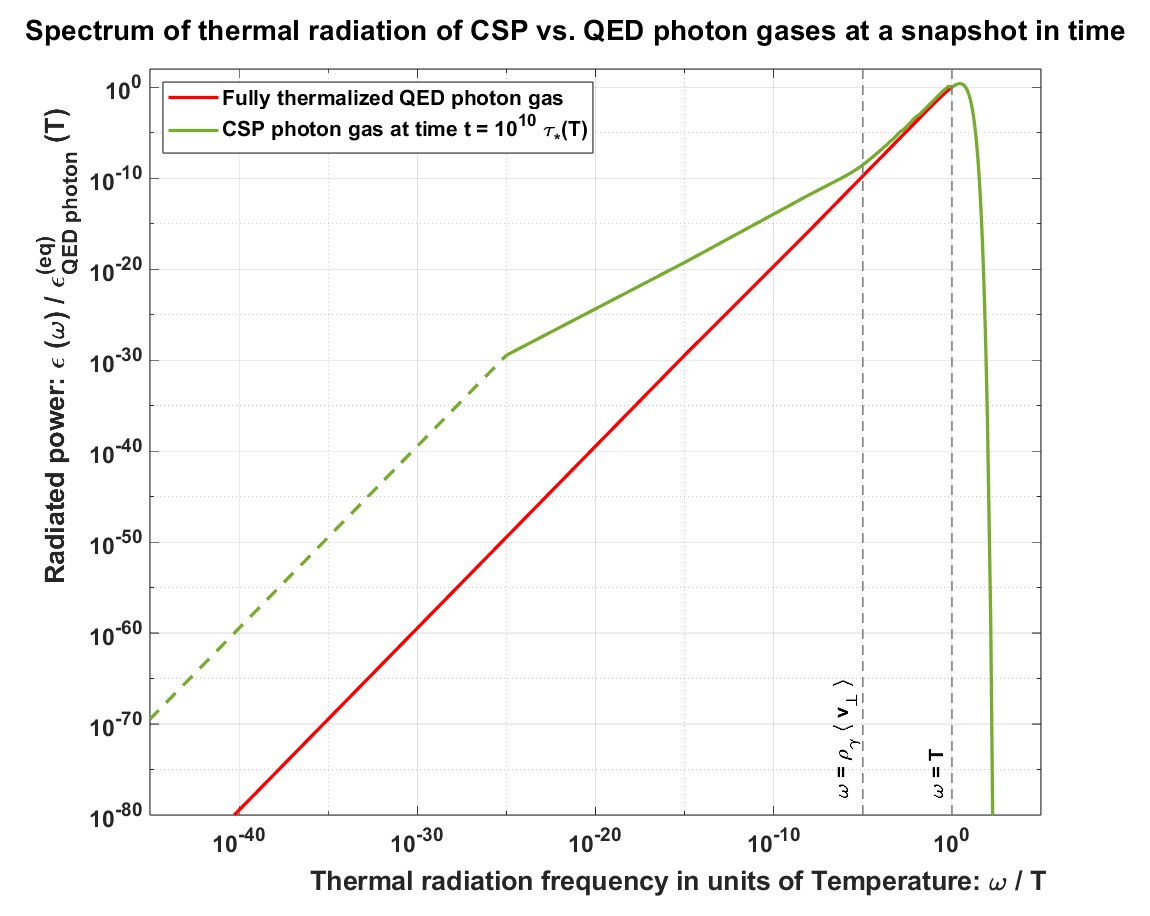}
    \caption{This figure illustrates the differences in the spectrum of thermal radiation of a CSP photon gas vs. a QED photon gas, at time $t = 10^{10} \tau_{*}(T)$, with discernible deviations from the familiar photon blackbody spectrum prominent at all deviation domain frequencies. We show a log-log plot to make the behavior across frequencies manifest. The x-axis is the radiated frequency $\omega$ in units of temperature $T$. The y-axis shows the power radiated at that frequency, in units of power radiated by a fully thermalized QED photon gas at the frequency $\omega = T$. The lowest frequency that has been populated to equilibrium density by any mode at this time is $\omega \approx 10^{-25} T$. The dashed portion of the green CSP line indicates an analytical extension to lower frequencies, where all energy radiated is from non-equilibrium contributions summed over all modes. Weaker deviations from the standard spectrum are also apparent at frequencies close to $T$ since the partner modes have populated some of the correspondence domain phase space at this time. Despite these strong deviations from standard spectrum in the deviation domain, the total energy radiated by the CSP photon gas is strongly dominated by frequencies higher than $\rog\vp$ - in this figure, the total energy radiated at frequencies above $\rog\vp$ is $\approx 10^{15} \times$ total energy radiated at frequencies below $\rog \vp$, a fact which might be obscured by the log-log scale. The parameter choices for this illustration are: Temperature $T = 10^{4} \rog$, $\vp = 0.1$ and time of snapshot $t = 10^{10} \tau_{*}(T)$.}
    \label{fig_Energy density as f (omega) at a given time}
\end{figure}

Figure \ref{fig_Energy density as f (omega) at a given time} illustrates all 3 aspects of CSP deviation domain behavior discussed above. 

\section{\label{Section_Synopsis and more}Synthesis: The effect of a non-zero spin scale on thermodynamic behavior}

So far, we have discussed the thermalization behavior of the CSP photon in the two energy regimes ($E \gg \rog \vp$ and $E \leq \rog \vp$), examining the key aspects of the behavior and building intuition step by step. In this section, we bring together a synthesis of the key results already discussed and present the complete picture of CSP thermalization behavior across all energies and times. We supplement this synthesis with some additional salient aspects of CSP thermodynamics whose discussion requires the complete picture. 

\subsection{Overall mode thermalization behavior}\label{Subsection_mode thermalization synopsis}

This subsection first brings together the key aspects of  thermalization behavior of a single mode across phase space. Subsequently, we synthesize all the helicity-based thermalization hierarchies. 

\subsubsection{The complete picture of mode thermalization}

Working in the soft limit, we saw that the thermalization behavior of all CSP modes follows the same pattern: Each mode has a characteristic energy, given by (\ref{eq_characteristic energy of a mode}), at which it its thermalization time is the shortest. As we move along the energy scale in either direction and away from the characteristic energy, the characteristic thermalization time of the mode increases, with $\tau_{h} \rightarrow \infty$ as $\eh \rightarrow 0$ and $\eh \rightarrow \infty$. At energies much greater than its characteristic energy, a mode's characteristic thermalization time grows with energy, following (\ref{eq_thermalization time ratio: E>rho}). At energies much less than its characteristic energy, the mode's characteristic thermalization time is inverse in energy, following the low helicity case in (\ref{eq_thermalization time ratio: E<<rho}). Notably, this behavior is also followed by the primary modes, whose characteristic energy is $\sim \rog\vp$. 

\begin{figure}
    \includegraphics[width= \columnwidth]{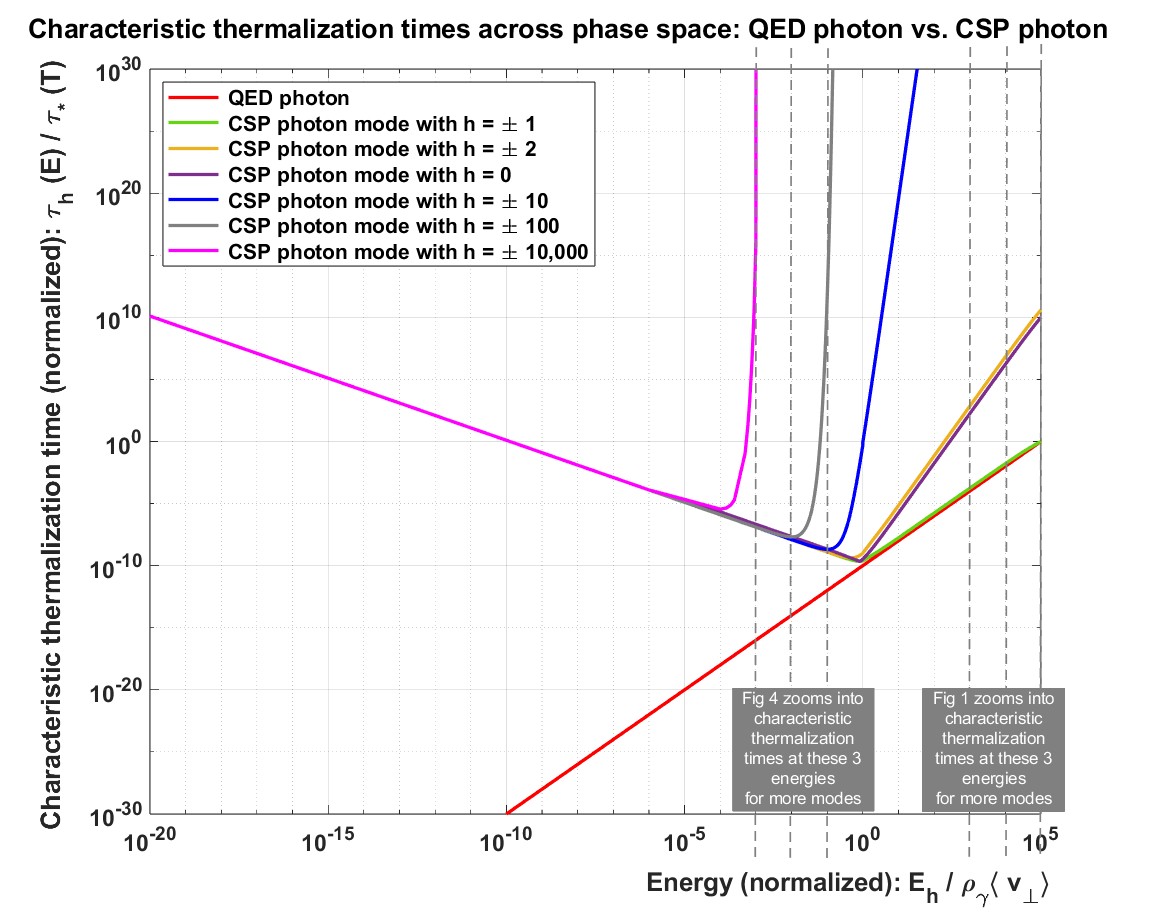}
    \caption{This log-log plot illustrates that for every thermalizing CSP photon mode, the characteristic thermalization times diverge in the UV and IR (but approach these divergences at different rates), keeping the total energy in the CSP photon gas well controlled at all times. The x-axis is shown in units of $\rog\vp$, and the y-axis is shown in units of benchmark thermalization time $\tau_{*}(T)$. We illustrate the behavior for primary modes, partner modes, and select high helicities chosen for illustrative purposes. It can be seen that: a) All modes have a characteristic energy $\echar$ given by (\ref{eq_characteristic energy of a mode}), at which it thermalizes fastest (given by (\ref{eq_thermalization time at characteristic energy})) - these can be read off at the minima for each mode in the figure b) Higher helicities have lower characteristic energies, and take increasingly longer to populate their phase space even at these characteristic energies. Note that $h = \pm 10,000$ modes have already increased their $\tau_{h}(E)/\tau_{*}(T)$ to $\mathcal{O}(10^{104})$ (beyond the y-axis cutoff in figure) at less than one order of magnitude in energy above their characteristic energy. The dashed lines denote the specific energies we used to illustrate mode thermalization hierarchies in Figure \ref{fig_thermalization times hierarchy: E>rho} and Figure \ref{fig_thermalization times: E<<rho}. In this figure, we include the hierarchy across energies to complete the picture of ``double hierarchy" in UV thermalization and ``weak hierarchy" in IR thermalization. The parameter choices for this illustration are: Temperature $T = 10^{4} \rog$, and $\vp = 0.1$.}
    \label{fig_mode thermalization time vs. energy}
\end{figure}

Figure \ref{fig_mode thermalization time vs. energy} illustrates the behavior of the characteristic thermalization time across phase space for select modes. 

This thermalization behavior means that each mode achieves equilibrium number density at its characteristic energy first. At any given time $t$, we can identify the lowest energy $\ehmin(t)$, and the highest energy $\ehmax(t)$, between which a mode has populated its equilibrium phase space. At times $t \ll \tau_{h}(\echar)$, $\ehmax(t)=\ehmin(t)=\echar$, which is as yet not fully populated. As time evolves above $\tau_{h}(\echar)$, $\ehmax(t)$ is given by (\ref{eq_Ehmax time dependence: E>rho}) and $\ehmin(t)$ by (\ref{eq_ehmin(t)}). Putting these together, the complete equation for the differential mode number density at any time is given by:

\begin{align} \label{eq_mode number density: all energies and times}
    n_{h}(\eh,t) \lessapprox 
     \begin{cases}
         \text{At ~time~}t \ll \tau_{h}\bigl(\echar\bigr):\\
         n_{h}^{(eq)}(\eh)\frac{t}{\tau_{h}(\eh)} & \text{for all}~~ \eh\\
         \\
          \text{At ~time~}t \geq \tau_{h}\bigl(\echar\bigr):\\
          n_{h}^{(eq)}(\eh)\frac{\eh}{\ehmin(t)} &0 \leq \eh \leq \ehmin(t)\\
            n_{h}^{(eq)}(\eh)  &\ehmin(t) \leq \eh \leq \ehmax(t) \\
            n_{h}^{(eq)}(\eh) (\frac{\ehmax(t)}{\eh})^{2(\Tilde{h}+1)} &\ehmax(t) \leq \eh < \infty \\ 
     \end{cases}
\end{align} 

\subsubsection{The complete picture of mode thermalization hierarchies}

While each mode follows this behavior, there are multiple hierarchies in the thermalization across modes that govern how the CSP gas as a whole behaves. The following hierarchies were discussed in \ref{Section:E>rho} and \ref{Section_E<<rho}:

\begin{enumerate}[(i)]
   \item The mode characteristic energy, given by (\ref{eq_characteristic energy of a mode}), is monotonic in $\Tilde{h}$, with $\echar \rightarrow 0$ for $\Tilde{h} \rightarrow \infty$ and $\echar \rightarrow \rog\vp$ for $\Tilde{h} \rightarrow 0$. This means modes with higher helicities first populate their phase space at smaller energies.  
    \item The mode thermalization time at its characteristic energy, given by (\ref{eq_thermalization time at characteristic energy}), which gives the shortest time taken by a mode to achieve equilibrium density at any region in its phase space, is also monotonic in $\Tilde{h}$, with $\tau_{h}\bigl(\echar\bigr) \rightarrow \infty$ as $\Tilde{h} \rightarrow \infty$. This means modes with higher helicities take parametrically longer to even populate the region of phase space close to their (already smaller) characteristic energies. 
    \item  In the correspondence domain, all modes have a ``double hierarchy" in their characteristic thermalization times relative to the primary mode per (\ref{eq_thermalization time ratio: E>rho}), due to cross sections suppressed via $\biggl(\frac{\eh}{\rog\vp}\biggr)^{2\Tilde{h}}$ as well as $2^{\Tilde{h}} (\Tilde{h}+1)!^{2}$. That is, higher helicities find it super-exponentially difficult to populate any region of their phase space, with the difficulty also increasing polynomially with energy. This hierarchy is reflected in the behavior of $\ehmax(t)$, which grows as $t^{\frac{1}{2(\Tilde{h}+1)}}$ per (\ref{eq_Ehmax time dependence: E>rho}).
    \item In the deviation domain, characteristic thermalization times follow a much weaker hierarchy, given by (\ref{eq_thermalization time ratio: E<<rho}).
\end{enumerate}

The mode thermalization hierarchies at energies much greater than and much less than the spin scale ((iii) and (iv) above) are illustrated in Figures \ref{fig_thermalization times hierarchy: E>rho} and \ref{fig_thermalization times: E<<rho} respectively. The hierarchies in mode characteristic energies and in the mode thermalization times at characteristic energy ((i) and (ii) above) are evident in Figure \ref{fig_mode thermalization time vs. energy}.

\subsection{Overall CSP behavior} \label{Subsection_CSP thermalization synopsis}

We first present the analyses estimating the effective relativistic degrees of freedom $g_{CSP}(t)$. We then synthesize the salient aspects of CSP thermalization behavior, energy density and spectrum of thermal radiation discussed previously across sections \ref{Subsection_number density: E>rho}, \ref{Subsection_Energy density: E>rho} and \ref{subsection_number & energy density: E<rho}.

\subsubsection{Effective relativistic degrees of freedom}

The total number density of the CSP gas is given by the sum over mode densities: 
\begin{align} \label{eq_CSP total number density integral}
     n_{CSP}(t) &= \sum_{h} \int_0^{\infty} d\eh n_{h}(\eh,t) \\
    &\equiv g_{CSP}(t) \frac{\zeta(3)}{\pi^{2}}T^{3} \label{eq_CSP total number density with g_CSP}
\end{align}
where we separate the familiar form for a thermalized Bose gas \cite{kolb_and_turner} to define the effective internal degrees of freedom for the CSP gas. $g_{CSP}(t)$ accounts for the fractional thermalization of every CSP mode and is a useful state variable of this system. Since the CSP gas is always partially thermal, it should be immediately apparent that $g_{CSP}(t) \lll$ the naive guess $\infty$, at any finite time. Since $g_{CSP}(t)$ depends on how much of its phase space each mode has populated by the time $t$, it implicitly depends on $\ehmax(t)$ and $\ehmin(t)$ of every CSP mode. 

Given the hierarchical, controlled thermalization behavior of the CSP at all energies, we can model the CSP as the familiar photon, but receiving time-dependent corrections to its relativistic degrees of freedom:
\begin{equation} \label{eq_g_CSP in terms of delta g(t)}
  g_{CSP}(t) \approx [2 - \Delta(t)] + \delta g(t)
\end{equation}
where the $2-\Delta(t)$ degrees of freedom correspond to the primary modes, and $\delta g(t)$ accounts for the correction from partial thermalization of all other modes. We estimate these effects using (\ref{eq_mode number density: all energies and times}): 
\begin{equation} \label{eq_gCSP(t)}
   \begin{split}
       g_{CSP}(t) \approx \sum_{h} \frac{1}{\zeta(3)} &\biggl[ \mathrm{Li}_{3}(e^{-\frac{\ehmin(t)}{T}})\\ &- \mathrm{Li}_{3}(e^{-\frac{\ehmax(t)}{T}})\\
       &+ \frac{\ehmin(t)}{T}\mathrm{Li}_{2}(e^{-\frac{\ehmin(t)}{T}})\\ &- \frac{\ehmax(t)}{T}\mathrm{Li}_{2}(e^{-\frac{\ehmax(t)}{T}})\\
       &+ \frac{\ehmin(t)^{2}}{2T^{2}}\mathrm{Li}_{1}(e^{-\frac{\ehmin(t)}{T}})\\ &- \frac{\ehmax(t)^{2}}{2T^{2}}\mathrm{Li}_{1}(e^{-\frac{\ehmax(t)}{T}}) \biggr]\\
   \end{split} 
\end{equation}

Comparing with (\ref{eq_g_CSP in terms of delta g(t)}) and noting that the primary modes will thermalize fastest, we recognize that at times $t \gg \tau_{*}(T)$, the two primary modes will contribute 2 full degrees of freedom, but with very small deviations due to their $\ehmin(t) > 0$ at all finite times:
\begin{equation} \label{eq_primary mode's very small deviation from 2 dofs}
    \begin{split}
        g_{\pm1}(t) \equiv [2 - \Delta(t)] \approx \frac{2}{\zeta(3)} \biggl[&\mathrm{Li}_{3}(e^{-\frac{\ehmin(t)}{T}}) \\&+ \frac{\ehmin(t)}{T}\mathrm{Li}_{2}(e^{-\frac{\ehmin(t)}{T}}) \\ &+ \frac{\ehmin(t)^{2}}{2T^{2}}\mathrm{Li}_{1}(e^{-\frac{\ehmin(t)}{T}}) \biggr]\\
    \end{split}
\end{equation}

Note that time- and $\rog$- dependence in (\ref{eq_gCSP(t)}) and (\ref{eq_primary mode's very small deviation from 2 dofs}) can be made explicit by using the equations for $\ehmin(t)$ in (\ref{eq_ehmin(t)}) and $\ehmax(t)$ in (\ref{eq_Ehmax time dependence: E>rho}). As $\ehmin(t) \rightarrow 0$ and $\ehmax(t) \rightarrow \infty$, a full degree of freedom is unlocked from a mode. We will consider these conditions as met when $\ehmin(t) \ll T$ and $\ehmax(t) \gg T$. 

For familiar thermodynamic systems that have $T \gg \rog$, $\ehmin(t)$ is always $\ll T$, so the behavior of the relativistic degrees of freedom tracks the evolution of $\ehmax(t)$ only. Specifically, it is driven by the strong mode thermalization hierarchies at energies much greater than the spin scale. For such thermal systems, the primary modes will be contributing 2 full degrees of freedom at all $t \gg \tau_{*}(T)$, and any deviations from this are negligible until the partner modes start appreciably thermalizing several time orders of magnitude after $\tau_{*}(T)$.  
Ultra cold thermal systems (that have $T \sim \rog\vp$) are likely to see detectable deviations in $g_{CSP}(t)$ and energy density in short times, and require further investigation.

\subsubsection{Synopsis of CSP thermalization behavior, \\ energy density and spectrum of thermal radiation}

The essence of the effects of a small non-zero $\rog$ on photon thermalization is two-fold: 
\begin{enumerate}[(i)]
    \item At energies above the spin scale (``correspondence domain"), we recover familiar thermodynamic behavior with $\rog$-dependent corrections to all thermodynamic quantities that become manifest as the system is allowed to thermalize over long periods of time. Specifically, in an isothermal system, we get corrections to internal energy, thermal power spectrum and relativistic degrees of freedom that are tell-tale signs of a CSP, but such deviations are observable only with exponentially long thermalization time scales. All thermodynamically relevant quantities remain finite at all finite times, and the rate of increase in CSP energy density is inverse in time. 
    \item We unlock an entirely new region of phase space at energies less than the spin scale (``deviation domain"), with novel behavior from all helicities. The CSP gas is populated nearly identically by increasing number of modes at progressively lower energies, but with thermalization timescales that also increase in tandem, keeping the IR phase space of the CSP gas well-behaved at all finite times. The total energy density in the deviation domain remains sub-dominant, but the power radiated at these very low frequencies shows large fractional deviations from QED, with calculable deviations in spectrum shape as well. 
\end{enumerate}

Since spin scale has mass dimensions, a thermodynamic system with non-zero $\rog$ has a new natural scale (in addition to temperature). The interplay of these two energy scales sets the overall thermal behavior of the CSP gas. Specifically, for a system with $T \gg \rog$, the dominant behavior of the CSP gas follows (i) above, whereas for a system with $T \sim \rog\vp$, we get entirely novel behavior overall, dominated by (ii) above. 

Since the spin scale is expected to be small in nature, the thermalization behavior of the CSP photon is identical to that of the ordinary photon for most familiar thermodynamic systems (which will have $T \gg \rog$). In such systems, any deviations in thermodynamic behavior due to the familiar photon potentially having a non-zero spin scale will be manifest only with very long thermalization timescales (assuming $\rog$ non-zero but not so small that it evades detectability in the age of the universe) and/or in the deviation domain behavior (assuming $\rog$ large enough to be detectable with available low energy technologies). 

Finally, we comment on the generality of the treatment in this paper based on the scaling of soft factors \eqref{eq_formfactor}.  There are multiple ways this might break down: First, the soft-factor decomposition \eqref{eq_amplitude split with soft limit} only captures the leading behavior of amplitudes when the kinematic invariants $k.p_i$ appearing in radiation from external legs are smaller than other kinematic invariants in the problem.  In renormalizable theories such as QED, this approximation introduces only ${\cal O}(1)$ errors to more energetic radiation. In any case, the most interesting phenomenology in CSP thermodynamics arises at CSP energies $E\ll T$, which is in the kinematic regime where \eqref{eq_amplitude split with soft limit} is applicable. 
A second objection to \eqref{eq_amplitude split with soft limit} is that its application to 2-to-2 processes, such as Compton scattering, is ill-defined because there is no associated ``parent'' amplitude $|{\cal M}|_0$ at real momenta.  However, CSP Compton scattering was explicitly studied in \cite{Schuster:2023jgc}, and found to have the same parametric $\rog$- and energy-scaling one would infer from \eqref{eq_amplitude split with soft limit} and \eqref{eq_formfactor}.  
A third and final concern is that our calculations were based on analysis of one soft factor associated with a single leg of an amplitude, rather than the sum over all soft factors in \eqref{eq_amplitude split with soft limit}, effectively neglecting interference between these emissions. This summation over legs is essential to the covariance of amplitudes obtained from \eqref{eq_amplitude split with soft limit}, and can give rise to physically important cancellations (for example, the amplitude for soft photon bremsstrahlung at frequency $\omega$ off an electron of initial velocity $\mathbf{v}$ that scatters by an amount $\mathbf{\Delta v} \ll v$ scales as $\mathbf{\Delta v}/\omega$, whereas our single-leg treatment would imply $\mathbf{v}/\omega$ scaling).  However, it is easy to check that partner-mode emissions do not experience analogous cancellations. Instead, the interfering soft factors from different legs have essentially random phases.  Thus, including interference effects should not appreciably change our conclusions.
Taking all of these effects into account, the thermalization times presented here should be viewed as parametric estimates, but the overall conclusions, including summations over all helicity modes, follow robustly from this parametrics.  

\section{Open Questions} \label{Section_open questions}

In this section, we briefly discuss the open questions that arise from the thermodynamic study of CSP photons presented in this paper. Resolving these is beyond the scope of this work, but could inspire aspects of our future study. 

We expect the spin scale of the CSP photon $\rog$ to be small in nature. If this weren't the case, the CSP photon gas would rapidly thermalize its partner and sub-partner modes, with rapid increase in its internal energy and degrees of freedom - essentially acting as a system with a high heat capacity that grows discernibly with time. In an isothermal bath, such as the one in early universe shortly before recombination, a photon gas with $\rho_{\gamma}$ much larger than $meV$-scale would have exhibited a fundamentally different behavior from what has been observed. For $\rho_{\gamma}$ smaller than $meV$-scale, we expect only small departures from standard thermal behavior, but it would be interesting to study how best to use early universe thermal signatures to constrain the spin-scale of the photon. 

In this paper, we investigated an isothermal system, with unbounded energy available to thermalize the CSP gas. Even with such a set up, we saw that the CSP gas takes infinite time to access that energy. Despite possessing infinite internal degrees of freedom in principle, a thermodynamic system with a CSP photon gas cannot access those degrees of freedom in any finite time - not even when supplied with infinite energy to do so. Now, let us consider a more physically realistic thermodynamic system. If we relax the isothermal assumption and supply fixed energy to the CSP gas, for instance. The primary modes of the CSP photon gas will still thermalize rapidly. The other modes still thermalize on exponentially longer time scales, but we can expect that some of the energy increase in modes that thermalize later will come from leakage of energy from the already thermalized modes. In such thermodynamic systems, the total energy in the CSP gas will be finite and bounded by the available energy, even with infinite time. We can expect the gas to slowly lower its overall temperature and increase its entropy as more modes thermalize. CSP thermalization behavior in this and other thermodynamic situations requires additional investigation.

Finally, ultra-cold thermal systems (those that have $T \lesssim \rog\vp$) would provide a potentially interesting regime in which to study signals of non-zero $\rho_{\gamma}$. To study such systems completely, we need to include Bose condensation effects and work with full scattering amplitudes, not just soft limits. This motivates further study of CSP physics (in QED) at low temperatures in future work.

%% file: Appendices.tex
\newpage

\section{Phase space evolution of the thermalizing CSP photon gas}\label{Appendix_boltzmann derivation}

The microscopic evolution of the phase space distribution of every mode $h$ of the photon gas is governed by the Boltzmann equation \cite{kolb_and_turner}:
\begin{equation} \label{appendix eq_Boltzmann equation}
    \hat{\textbf{L}}[f_{h}] = \textbf{C}[f_{h}]
\end{equation}
where $\textbf{C}$ is the collision operator and $\hat{\textbf{L}}$ is the Liouville operator. The covariant, relativistic Liouville operator is \cite{kolb_and_turner}:
\begin{equation}
    \hat{\textbf{L}} = p^{\mu}\frac{\partial}{\partial x^{\mu}} - \Gamma^{\mu}_{\nu \sigma} p^{\nu} p^{\sigma} \frac{\partial}{\partial p^{\mu}}
\end{equation}
In a Minkowski background, the Liouville operator simplifies to:
\begin{equation} \label{appendix eq_Liouville factor}
     \hat{\textbf{L}}[f_{h}(\eh,t)] = E \frac{\partial}{\partial t}f_{h} (\eh,t)
\end{equation}
The collision term for the process $a+b+... \longleftrightarrow i+j+\gamma_{h}+...$ is given by \cite{kolb_and_turner}:
\begin{align} \label{appendix eq_collision factor}
    \hat{\textbf{C}}[&f_{h}(\eh,t)] = \int d\Pi_{a} d\Pi_{b} ....~ d\Pi_{i} d\Pi_{j}.... \nonumber\\ &\times (2\pi)^{4} \delta^{4}(\Sigma p_{in}-\Sigma p_{out + \gamma_{h}}) \nonumber\\ &\times \bigg[|\mathcal{M}|^{2}_{a + b + .. \rightarrow i + j+ \gamma_{h}+..} f_{a} f_{b}... (1 \pm f_{i}) (1 \pm f_{j}) (1 \pm f_{h}).. \nonumber\\ & - |\mathcal{M}|^{2}_{i + j+ \gamma_{h}+ .. \rightarrow a+b+..} f_{i} f_{j} f_{h}.. (1 \pm f_{a}) (1 \pm f_{b}) .. \bigg]
\end{align}
where `in' denotes the incoming scatterers, `out' denotes all outgoing particles except the CSP mode $\gamma_{h}$ that we are interested in, $f_{\psi}$ refers to the phase space distribution function of particle $\psi$, $d\Pi \equiv \frac{g}{(2\pi)^{3}}\frac{d^{3}p}{2E}$, and $(1 \pm f_{\psi})$ factors capture Bose enhancement/ Fermi suppression respectively. We invoke time invariance to assume  
$|\mathcal{M}|^{2}_{a + b + .. \rightarrow i + j+ \gamma_{h}+..} =  |\mathcal{M}|^{2}_{i + j+ \gamma_{h}+ .. \rightarrow a+b+..} \equiv |\mathcal{M}|^{2}$. Using the energy conservation part of the delta function, and assuming chemical and kinetic equilibrium of all other species, (\ref{appendix eq_collision factor}) simplifies to:
\begin{equation} \label{appendix eq_simplified collision factor}
    \begin{split}
        \hat{\textbf{C}}[f_{h}& (\eh,t)] = \int d\Pi_{a} d\Pi_{b} ~...~ d\Pi_{i} d\Pi_{j} ~... \\& \times (2\pi)^{4} \delta^{4}(\Sigma p_{in}-\Sigma p_{out + \gamma_{h}})|\mathcal{M}|^{2} \\& \times f_{a} f_{b}... (1 \pm f_{i}) (1 \pm f_{j})...\bigg[ 1 + f_{h}[1- \exp{\frac{\eh}{T}}] \bigg]
    \end{split}
\end{equation}
where T is the temperature. Note that while all equations upto (\ref{appendix eq_collision factor}) are explicitly Lorentz covariant, (\ref{appendix eq_simplified collision factor}) is not, since we single out the frame of reference in which the temperature T is a monopole to do thermodynamics. This is the frame in which we will specify all particle distribution functions. 

Recognizing that in equilibrium, $f_{h}$ will  follow Bose-Einstein statistics, we can use $f_{h}^{(eq)}(\eh) = \frac{1}{\exp{\frac{\eh}{T}}-1}$ to rewrite the $\bigg[ 1 + f_{h}[1- \exp{\frac{\eh}{T}}] \bigg]$ factor in (\ref{appendix eq_simplified collision factor}) as $\bigg[ 1 - \frac{f_{h}(\eh,t)}{f_{h}^{(eq)}(\eh)} \bigg]$. Using (\ref{appendix eq_simplified collision factor}) and (\ref{appendix eq_Liouville factor}), the Boltzmann equation (\ref{appendix eq_Boltzmann equation}) reduces to a differential equation that can be solved under the assumptions already stated, to give:
\begin{equation}
    f_{h}(\eh,t)= f_{h}^{(eq)}(\eh)[1-\exp{(-t/\tau_{h}(\eh))}] 
\end{equation} 
where $\tau_{h}(\eh)$ can be recognized as:
\begin{equation}
\begin{split}
    \tau_{h}(\eh)=&f_{h}^{(eq)}(\eh) \bigg[\int d\Pi_{in}f_{in} (1 \pm f_{i}) (1 \pm f_{j})... d\Pi_{out}\\ & \times (2\pi)^{4}\delta^{4}(\Sigma p_{in}-\Sigma p_{out+\gamma_{h}})  |\mathcal{M}|^{2}\frac{1}{\eh}\bigg]^{-1}
\end{split}
\end{equation}
Ignoring the Bose enhancement/ Fermi suppression factors from the outgoing states, i.e., taking $(1 \pm f_{\psi}) \approx 1$, we get (\ref{eq_thermalization time}). When we consider a multi-photon scattering process, this assumption means that we ignore any Bose enhancement to the production of partner and sub-partner modes from the faster thermalization of primary modes. We make the reasonable assumption that multi-photon CSP processes are sub-dominant to single photon scattering processes, as in familiar QED. Additionally, familiar QED has the same IR divergence from Bose-Einstein statistics - so this aspect is not unique to CSP photons. Hence, we find it best to keep the Bose enhancement aside and focus only on aspects of thermal behavior arising from a non-zero spin scale. 

\section{Modifications for a non-relativistic thermal scatterer}\label{Appendix_delta}

In the main paper, we used the average velocity $\vp$ in equations, and used Taylor expansion and/or asymptotic forms of $J_{h}(\frac{\rog v_{\perp}}{E})$ where it was valid to do so. These simplifications were made mainly to aid intuitive understanding of CSP behavior. This appendix provides some details on modifications needed to some of these simplified equations when the scatterer is non-relativistic. Note that all numerical simulations presented in the paper used the full thermal distribution of $v_{\perp}$ as well as the full Bessel function form of the soft limit scattering amplitudes (without any simplifications). 

Since scattering cross sections have a velocity dependence in the soft limit per (\ref{eq_amplitude split with soft limit}), (\ref{eq_z}) and (\ref{eq_formfactor}), the velocity distribution of a non-relativistic scatterer needs to be accounted for when calculating the mode thermalization times. Equation (\ref{eq_thermalization time}) includes calculations with the following form, using a Boltzmann distribution for the non-relativistic scatterer:
\begin{equation} \label{appendix eq_thermal integral}
    \tau_{h}(E) \supset \int_{0}^{1} d v_{\perp} v_{\perp} \exp \biggl({-\frac{v_{\perp}^{2}}{2\vp^{2}}}\biggr) ~ \bigg| J_{h}(\frac{\rog v_{\perp}}{E}) \bigg| ^{2}
\end{equation}
We can approximate $J_{\alpha}(x)$ with the first term in its Taylor expansion \cite[\href{http://dlmf.nist.gov/10.2.E2}{(10.2.2)}]{NIST:DLMF} when:
\begin{align}
    J_{\alpha}(x) &= \sum_{m=0}^{\infty} \frac{(-1)^{m} (\frac{x}{2})^{2m+\alpha}}{m! ~\Gamma (m + \alpha + 1)} \label{appendix eq_Bessel Taylor expansion}\\
    & \approx \frac{(\frac{x}{2})^\alpha}{\Gamma (\alpha + 1)} ~~\text{when}~ {x \ll 2 \sqrt{\alpha + 1}} \label{appendix eq_Bessel Taylor validity}
\end{align}
When the simplifying condition in (\ref{appendix eq_Bessel Taylor validity}) is valid for the entire range of the integration in (\ref{appendix eq_thermal integral}), the thermal averaging will give a lower incomplete gamma function $\gamma (\Tilde{h}+1,\frac{1}{2\vp^{2}})$, which is $\approx (\Tilde{h}+1)!$ only if $ \Tilde{h}+1 < \frac{1}{2\vp^{2}}$. This means that when working with the simplified forms of the Bessel scattering cross sections, we need to be careful about the range of validity of the simplifications. The rest of this appendix discusses the modifications that need to be made for certain equations in the main paper to account for the effect of thermal distribution of non-relativistic scatterer velocities.

Equation (\ref{eq_thermalization time ratio: E>rho}) gets modified as: 
    \begin{equation} \label{appendix eq_thermalization time ratio: E>rho}
        \frac{\tau_{h}(E)}{\tau_{*}(E)} \sim \frac{\bigl\langle|z|^{2}\bigr\rangle}{\bigl\langle|z F_{h}(\rog z)|^{2}\bigr\rangle} \; \approx \; 2^{\Tilde{h}} (\Tilde{h}+1)!^{(2-\delta)} \biggl(\frac{\eh}{\rog\vp}\biggr)^{2\Tilde{h}}
    \end{equation}
where the continuous parameter $\delta$ accounts for the weaker suppression of smaller helicities due to the thermal distribution in scatterer velocities when we consider a non-relativistic scatterer. $\delta=0$ for all helicities when the scatterer is relativistic. When we consider non-relativistic scatterers, $0 \leq \delta \leq 1$. Smaller helicities will see a shorter time vs. that obtained using $\vp$ in lieu of a full thermal distribution of $v_{\perp}$ i.e., $\delta \rightarrow 1$ for only those modes with $\Tilde{h} < \frac{1}{2\vp^{2}}$.

Equation (\ref{eq_d/dt energy density with T/E assumption}) requires a minor modification with $(\Tilde{h}+1)!^{2}$ in the denominator replaced with $\Tilde{h}+1)!^{(2-\delta)}$. This manifests as a change in the helicity scaling of $\frac{d}{dt} \ed_{CSP}(t)$, which we discussed in the main paper as following $\Tilde{h}^{-4}$, where 3 powers of $\Tilde{h}$ came from $(\Tilde{h}+1)!^{\frac{3}{(\Tilde{h}+1)}}$ and 1 came directly from the $2 \Tilde{h}$ in the denominator. Including the $\delta$ parameter, the helicity scaling varies as $\Tilde{h}^{2.5-4}$ for a non-relativistic scatterer. Low helicities in the non-relativistic limit have $\delta \approx 1$ as above and $(\Tilde{h}+1)!^{\frac{1.5}{(\Tilde{h}+1)}}$ grows as $(\Tilde{h}+1)^{1.5}$. High helicities have $\delta = 0$ as explained above and $(\Tilde{h}+1)!^{\frac{3}{(\Tilde{h}+1)}}$ grows as $(\Tilde{h}+1)^{3}$. In the relativistic limit, all modes get suppressed with $(\Tilde{h}+1)^{3}$. These modifications do not change anything significant about $\ed_{CSP}(t)$ since its behavior is controlled by other aspects of (\ref{eq_d/dt energy density with T/E assumption}). The only key property of $\ed_{CSP}(t)$ that depended on the helicity scaling is the sum convergence, and this continues to hold with the modified scaling for low helicities. 

Equation (\ref{eq_ mode energy density as function of time}), which directly followed from (\ref{eq_d/dt energy density with T/E assumption}), gets the same modifications discussed in the previous paragraph. For a relativistic scatterer, the power law dependence for all modes has $(\Tilde{h}+1)^{3}$ in the denominator. For a non-relativistic scatterer, this gets modified to be $(\Tilde{h}+1)$ for low helicities with $\Tilde{h} < \frac{1}{2\vp^{2}}$ and $(\Tilde{h}+1)^{3}$ for higher helicities. 

Note that equation (\ref{eq_thermalization time ratio: E<<rho}), which is valid for $\eh \ll \rog\vp$ already implicitly accounted for the ranges in $\delta$.  

\section{Time evolution of CSP energy density}\label{Appendix_change in energy density}

The energy density in the CSP gas can be obtained using \cite{kolb_and_turner}:

\begin{equation} \label{appendix eq_CSP energy density}
   \ed_{CSP}(t) = \sum_{h} \int_0^{\infty} d\eh n_{h}(\eh,t).\eh 
\end{equation}

Using (\ref{eq_mode number density: all energies and times}), we can write (\ref{appendix eq_CSP energy density}) as:

\begin{equation} \label{appendix eq_CSP energy density detail-1}
\begin{split}
        \ed_{CSP}(t) \lessapprox \sum_{h} \bigg[&\int_0^{\ehmax(t)} d\eh n_{h}^{(eq)} (\eh)\eh \\+ &\int_{\ehmax(t)}^{\infty} d\eh n_{h}^{(eq)}(\eh)\eh \frac{t}{\tau_{h}(\eh)}\bigg]
\end{split}
\end{equation}

In writing (\ref{appendix eq_CSP energy density detail-1}), we have used the following: i) Overall energy density in any mode is most sensitive to the highest occupied energy, and comparatively insensitive to the lowest occupied energy, allowing us to take $\ehmin(t) \rightarrow 0$ ii) $\ehmax(t)$ tracks (\ref{eq_Ehmax time dependence: E>rho}), with characteristic thermalization time tracking (\ref{eq_thermalization time ratio: E>rho}) since $\ehmax(t) \geq \echar$ always.

Taking the time derivative of (\ref{appendix eq_CSP energy density detail-1}), and noting that $n_{h}(\eh)$ is continuous at all energies including $\ehmax(t)$, we get:

\begin{equation} \label{appendix eq_CSP energy density time change}
   \frac{d}{dt}\ed_{CSP}(t) \lessapprox \sum_{h} \int_{\ehmax(t)}^{\infty} d\eh n_{h}^{(eq)}(\eh) \frac{\eh}{\tau_{h}(\eh)}
\end{equation}

To make (\ref{appendix eq_CSP energy density time change}) tractable for further study, we approximate $f_{h}^{(eq)}(\eh)$ as $\frac{T}{\eh} ~~\text{for all} ~\eh \leq T$ and as $\exp (-\frac{\eh}{T}) ~~\text{for all}~ \eh > T$. We study the two regimes separately, and write:

\begin{equation} \label{appendix eq_CSP energy density time change detail -1}
\begin{split}
    \frac{d}{dt}\ed_{CSP}(t) \lesssim \frac{1}{2\pi^{2}} \bigg[&\sum\limits_{h \, : \, \substack{\tau_{h}(T) \geq t \\ h \neq \pm1}} \int_{\ehmax(t)}^{\infty} d\eh \frac{T \eh^{2}}{\tau_{h}(\eh)} \\+ &\sum\limits_{h \, : \, \substack{\tau_{h}(T) < t \\ h \neq \pm1}} \int_{\ehmax(t)}^{\infty} d\eh e^{-\frac{\eh}{T}} \frac{\eh^{3}}{\tau_{h}(\eh)}\bigg]
\end{split}
\end{equation}

In (\ref{appendix eq_CSP energy density time change detail -1}), the former sum is taken over modes that are still thermalizing, given by the condition $\tau_{h}(T) \geq t$. The latter sum is taken over modes that have already thermalized their phase space upto energies $\eh > T$, given by the criterion $\tau_{h}(T) < t$. We exclude the primary modes from these sums since we assume them to be thermal in this analysis, and will use $\tau_{*}(T)$ as the benchmark time in the next step. Substituting (\ref{eq_thermalization time ratio: E>rho}) in (\ref{appendix eq_CSP energy density time change detail -1}), the integration in both sums can be exactly done.

\begin{equation} \label{appendix eq_CSP energy density time change detail -2}
    \begin{split}
        \frac{d}{dt} \ed_{CSP}(t) \lesssim &\sum\limits_{h \, : \, \substack{\tau_{h}(T) \geq t \\ h \neq \pm1}} \frac{T^{4}}{2\pi^{2}[2\Tilde{h}-1]} t^{-1} \left( \frac{\ehmax(t)}{T} \right)^{3} \\
        + &\sum\limits_{h \, : \, \substack{\tau_{h}(T) < t \\ h \neq \pm1}} \bigg [\frac{T^{4}}{2\pi^{2}} t^{-1} \left( \frac{\ehmax(t)}{T} \right)^{2(\Tilde{h}+1)} \\ 
        &~~~~~~~~~~~~~~~~~~~~\times \Gamma \big (2-2\Tilde{h},\frac{\ehmax(t)}{T} \big) \bigg ]
    \end{split}
\end{equation}

In (\ref{appendix eq_CSP energy density time change detail -2}), $\Gamma (x,y)$ is the upper incomplete gamma function. Since the second sum is taken over modes that have $\ehmax(t) > T$, the gamma function falls exponentially or faster with time for $\Tilde{h} \geq 1$. So, as long as there are at least some modes with $\tau_{h}(T) \geq t$ i.e., at all finite times, the dual sum in (\ref{appendix eq_CSP energy density time change detail -2}) is always dominated by the sum over thermalizing modes. The time dependence in (\ref{appendix eq_CSP energy density time change detail -2}) can be made explicit with (\ref{eq_Ehmax time dependence: E>rho}). Doing so, and dropping the sum over thermal modes that is sub-dominant gives the expression in (\ref{eq_d/dt energy density with T/E assumption}).